\documentclass[12pt]{article}
\usepackage{graphicx}
\usepackage{cite}
\usepackage{color}
\usepackage{amssymb}
\def\rbbfm{0.221}
\def\rnnfm{1.04}
\def\rssfm{0.47}
\def\mX{5.634}
\def\BBfacIi{2.28}
\def\BBfacIii{0.436}
\def\BBfacIiii{0.236}
\def\BBfacIiv{0.150}
\def\bivS{0.108}
\def\CivS{10.735}
\def\WivS{38}
\def\CiiiD{10.867}
\def\WiiiD{42}
\def\CvS{11.017}
\def\bareminusCvS{236} 
\def\WvS{59}
\def\Bzero{0.122}
\newcommand{\babar}{{\it BABAR}}
\newcommand{\bm}[1]{\mbox{\boldmath $#1$}}
\newcommand{\fnd}[2]{\frac{\textstyle #1}{\textstyle #2}}
\newcommand{\abs}[1]{\left| #1\right|}
\newcommand{\braket}[3]{\mbox{$\left\langle #1\left|
#2\right| #3\right\rangle$}}
\newcommand{\Imag}[1]{\Im {\it m}(#1 )}
\newcommand{\fndrs}[4]{\fnd{\raisebox{#1}{$#2$}}{\raisebox{#3}{$#4$}}}
\newcommand{\x}[1]{{\textstyle #1}}
\newcommand{\xrm}[1]{{\textstyle \mbox{\rm #1}}}

\newcommand{\lijntje}[3]{(\begin{picture}(20,0)(4,0)
\end{picture})}
\definecolor{pink}{rgb}{0.5,0.0,1.0}
\begin{document}
\title{Mass and width of the $\Upsilon (4S)$}
\author{
Eef~van~Beveren$^{\; 1}$ and George~Rupp$^{\; 2}$\\ [10pt]
$^{1}${\small\it Centro de F\'{\i}sica Computacional,
Departamento de F\'{\i}sica,}\\
{\small\it Universidade de Coimbra, P-3004-516 Coimbra, Portugal}\\
{\small\it eef@teor.fis.uc.pt}\\ [10pt]
$^{2}${\small\it Centro de F\'{\i}sica das Interac\c{c}\~{o}es Fundamentais,}\\
{\small\it Instituto Superior T\'{e}cnico, Universidade T\'{e}cnica de
Lisboa,}\\
{\small\it Edif\'{\i}cio Ci\^{e}ncia, P-1049-001 Lisboa, Portugal}\\
{\small\it george@ist.utl.pt}\\ [.3cm]
{\small PACS number(s): 14.40.Gx, 13.25.Gv, 14.40.Nd, 11.80.Gw}
}


\maketitle
\centerline{\scalebox{1.0}{\includegraphics{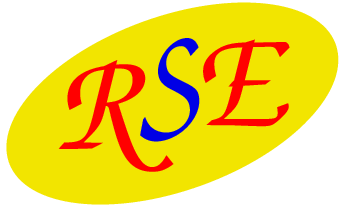}}}
\vspace{10pt}

\begin{center}
{\bf Collaboration for the Resonance-Spectrum Expansion}
\end{center}
\begin{abstract}
Recent data on $e^-e^+\to b\bar{b}$ by the \babar\ \/Collaboration
[B.~Aubert {\it et al.}, Phys.\ Rev.\ Lett.\ {\bf102}, 012001 (2009)]
in the energy range delimited by the $B\bar{B}$ and
$\Lambda_{b}\bar{\Lambda}_{b}$ thresholds are analyzed in a multichannel
formalism that incorporates the usual Breit-Wigner resonances, but interfering
with a background signal due to the opening of open-bottom thresholds.
In particular, the $\Upsilon (4S)$ resonance is determined to have a mass
of \CivS\ GeV and a width of \WivS\ MeV. Also two higher $\Upsilon$ resonances
are identified, parametrized, and classified.

Moreover, it is found that near the $B\bar{B}$ threshold
open-bottom production in electron-positron annihilation
is dominated by the reaction chain
$e^{-}e^{+}\to n\bar{n}\to B\bar{B}$ ($n=u/d$)
rather than 
$e^{-}e^{+}\to b\bar{b}\to B\bar{B}$,
whereas near the $B_{s}\bar{B}_{s}$ threshold
the reaction chain
$e^{-}e^{+}\to s\bar{s}\to B\bar{B}$ dominates
the production amplitude.

The vital role played in this analysis
by the {\it universal confinement frequency} \/$\omega$,
defined in 1980
[E.~van Beveren, C.~Dullemond, and G.~Rupp, Phys.\ Rev.\ D {\bf21}, 772 (1980)]
and accurately determined in 1983 
[E.~van Beveren, G.~Rupp, T.~A.~Rijken, and C.~Dullemond, 
Phys.\ Rev.\ D {\bf27}, 1527 (1983)],
is further confirmed.
\end{abstract}

\section{Introduction}

In recent years, several experimental groups,
in particular the \babar\ \/Collaboration,
have published data that were only partly analyzed,
thus offering us the opportunity to also interpret
measurements outside the invariant-mass regions of 
direct interest for these experiments.
We welcome this attitude, since it supplies us with valuable
pieces of information, allowing us to test our theoretical methods
on a larger data range.
For example, in data on the enhancement in the
$e^{+}e^{-}\to\Lambda_{c}^{+}\Lambda_{c}^{-}$ cross sections
reported by the Belle Collaboration \cite{PRL101p172001},
we could observe clearly two new charmonium resonances,
viz.\ the $\psi (5S)$ and $\psi (4D)$ \cite{ARXIV09080242},
as well as glimpses of what we believe to be the $\psi (6S)$ and
$\psi (5D)$ \cite{EPL85p61002} states. Furthermore,
in data on $e^+e^-\to J/\psi\pi\pi$
and the $X(4260)$ enhancement published by the \babar\ \/Collaboration
\cite{PRL95p142001,ARXIV08081543},
we observed
the $\psi (3D)$ resonance \cite{ARXIV09044351}.
Finally, we identified an interference phenomenon
between OZI-allowed and OZI-forbidden decay modes \cite{PRD79p111501R}
near the $X(4260)$ enhancement
in the \babar\ \/data of Ref.~\cite{ARXIV08081543}.

It may have come as a surprise to our experimental colleagues that 
their observed enhancements could be interpreted either as 
normal $q\bar{q}$ resonances or as threshold effects, whereas
sporadic coincidences with some of the countless states of the
tetraquark, hybrid or molecular spectra, predicted by current
theories and models of strong interactions, were not accompanied
by a simultaneous description of the production data.
\begin{figure}[htbp]
\begin{center}
\begin{tabular}{c}
\scalebox{1.0}{\includegraphics{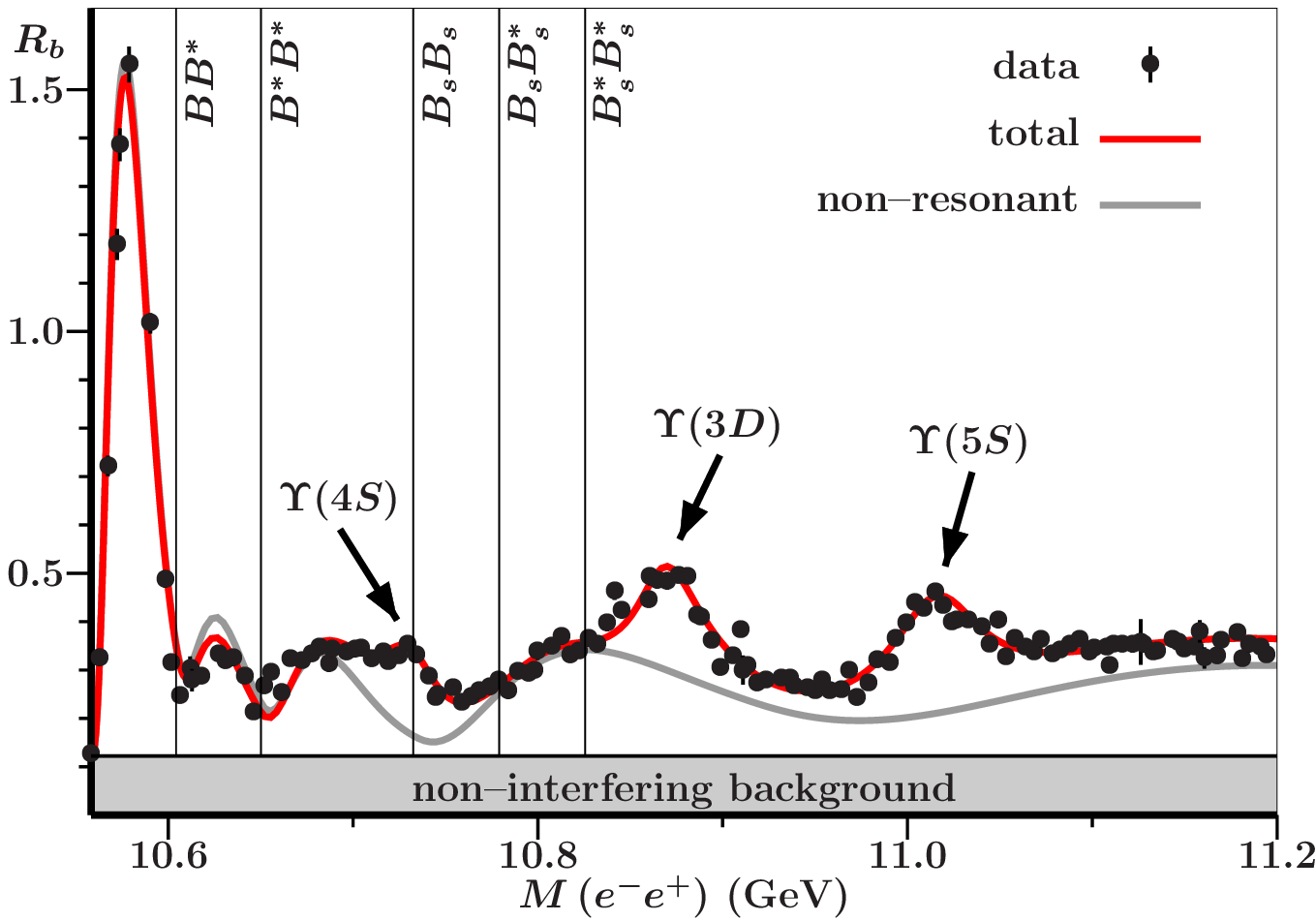}} \\
\end{tabular}\\ [-15pt]
\end{center}
\caption[]{\small $R_b$ as measured by \babar\ \/\cite{PRL102p012001}
in electron-positron annihilation for energies between
the $B\bar{B}$ and $\Lambda_{b}\bar{\Lambda}_{b}$ thresholds,
vs.\ our model fit \lijntje{1}{0}{0} as
discussed in Secs.~\ref{formalities} and \ref{parameters}.
The nonresonant contribution is also shown \lijntje{0.6}{0.6}{0.6}.
Thresholds of $B\bar{B}^{\ast}+\bar{B}B^{\ast}$,
$B^{\ast}\bar{B}^{\ast}$, $B_{s}\bar{B}_{s}$,
$B_{s}\bar{B}_{s}^{\ast}+\bar{B}_{s}B_{s}^{\ast}$
and $B_{s}^{\ast}\bar{B}_{s}^{\ast}$ are indicated,
as well as the central masses of the $\Upsilon (4S)$,
$\Upsilon (3D)$ and $\Upsilon (5S)$ resonances.
}
\label{babarups}
\end{figure}
If we compare this situation with the very good description
of experimental results for QED,
it seems fair to conclude that strong interactions
and, in particular, confinement are only poorly understood.
Actually, we conclude that this is the real challenge from experiment
to theory and models: to figure out what more is needed
in order to fully and in detail describe the experimental data, and not
just the central masses of some bumps.
In the present paper, we take up this challenge,
hoping that it may contribute to a further understanding
of the interactions between quarks and gluons,
and their relation to the observed structures
in scattering and production.
Our results for vector bottomonium are depicted in Fig.~\ref{babarups}
and will be discussed
in Secs.~\ref{formalities} and \ref{parameters}.

From combined data, published by the \babar\ \/Collaboration
in Refs.~\cite{PRD72p032005,PRL102p012001},
we concluded \cite{ARXIV09080242}
that for the enhancement just above the $B\bar{B}$ threshold
a description in terms of a wave function
with a dominant $B\bar{B}$ component appears to be more adequate
than assuming a pole in the scattering amplitude
due to a supposed underlying $b\bar{b}$ state.
Consequently, the enhancement does not represent
the $\Upsilon (4S)$ resonance.
We shall show that the true $\Upsilon (4S)$ 
lies about 160 MeV higher, viz.\ at \CivS\ GeV.

In the following, we study data on hadron production
in electron-positron annihilation
in the invariant-mass interval between
the $B\bar{B}$ and $\Lambda_{b}\bar{\Lambda}_{b}$ thresholds,
published by the \babar\ \/Collaboration \cite{PRL102p012001}.
This paper focuses on two of the resonances in the $b\bar{b}$ spectrum,
using data obtained with the \babar\ \/detector at the PEP-II storage ring,
resulting in the Breit-Wigner (BW) parameters
$10876\pm 2$~MeV (mass) and $43\pm 4$~MeV (width)
for the $\Upsilon (10860)$, and
$10996\pm 2$~MeV (mass) and $37\pm 3$~MeV (width)
for the $\Upsilon (11020)$.
These values differ substantially, in particular for the widths,
from earlier results
of the CUSB \cite{PRL54p377} and CLEO \cite{PRL54p381} Collaborations,
and also from the world averages \cite{PLB667p1}
$10865\pm 8$~MeV (mass) and $110\pm 13$~MeV (width)
for the $\Upsilon (10860)$, and
$11019\pm 8$~MeV (mass) and $79\pm 16$~MeV (width)
for the $\Upsilon (11020)$.
Obviously, such discrepancies call for further study.

In the present work, we apply a recently developed formalism
for the study of hadronic electron-positron annihilation reactions
to determine cross sections for open-bottom production
above the $B\bar{B}$ threshold.
The formalism is outlined in Sec.~\ref{formalities}
and further discussed in Sec.~\ref{parameters}.
Details of the results are presented in Sec.~\ref{Results}.
In Sec.~\ref{future} we propose future analyses
on existing data in order to further clarify some of our findings.
Conclusions are drawn in Sec.~\ref{finalities}.

\section{Open-bottom production in \bm{e^{-}e^{+}} annihilation}
\label{formalities}

In Ref.~\cite{AP323p1215}, we derived a relation between
the production amplitude $a_{\ell}(\alpha\to i)$
for $e^{+}e^{-}$ annihilation into two mesons
and the matrix elements of the meson-meson scattering amplitude $T$,
resulting for the $\ell$-th partial wave in an expression of the form
\begin{equation}
a_{\ell}(\alpha\to i)\propto
g_{\alpha i}\, j_{\ell}\left( p_{i}r_{0}\right)
+\frac{i}{2}\sum_{\nu}\,
h^{(1)}_{\ell}\left( p_{\nu}r_{0}\right)\,
g_{\alpha\nu}\,
T_{\ell}(\nu\to i)
\;\;\; ,
\label{prodamp}
\end{equation}
where the $g_{\alpha i}$ stand for the relative couplings of each of
the two-meson systems $i$ to a $q\bar{q}$ state of flavor $\alpha$,
and $j_{\ell}$ and $h^{(1)}_{\ell}$ are the spherical Bessel and 
Hankel function of the first kind, respectively.
The radius $r_{0}$ represents
the average distance at which the meson pair emerges from the interaction
region. Furthermore, $\vec{p}_{i}$ is
the relative linear momentum in two-meson channel $i$.
The matrix element $T_{\ell}(\nu\to i)$ of
the scattering amplitude $T$ describes transitions between
channels $\nu$ and $i$ in multichannel meson-meson scattering.

The generic form of the scattering amplitude is, in the
Resonance-Spectrum Expansion (RSE) \cite{NTTP4}, given by
(with $M=\sqrt{s}$)
\begin{equation}
T_{ij}(M)=\fndrs{2pt}{A_{ij}(M)}{-0pt}{D(M)}
\;\;\; ,
\label{generic}
\end{equation}
which satisfies the unitarity condition 
\begin{equation}
\Imag{D^{\ast}A_{ij}}=\sum_{\nu}\, A_{i\nu}A_{j\nu}
\;\;\; .
\label{Ucondition}
\end{equation}
We have shown in Ref.~\cite{AP323p1215}
that relation (\ref{prodamp}) satisfies
the extended unitarity theorem of Watson \cite{PR88p1163}.
The denominator $D$ of expression (\ref{generic})
is in the RSE given by the expansion
\begin{equation}
D(M)=1+2i\sum_{n,\nu}\,\fnd{g_{n\nu}^{2}\, G_{\nu}(M)}{M-E_{n}}
\;\;\; .
\label{denominator}
\end{equation}
The sum in Eq.~(\ref{denominator})
runs over the whole confinement spectrum ($n=0,\ldots,\infty$),
with levels given by $E_{n}$,
and over all meson-meson channels involved, labeled by $\nu$,
the kinematics of which is contained in the functions $G_{\nu}(M)$.
The relative couplings of the two-meson channels to
the states of the confinement spectrum are given by $g_{n\nu}$
\cite{ZPC21p291}.
In Ref.~\cite{AP324p1620}, we showed that a denominator
of the form given in Eq.~(\ref{denominator})
is equivalent to an effective meson theory with an infinite
number of seeds, albeit with a well-defined mass spectrum.

The poles of the scattering amplitude can be searched for
by solving the equation $D(M)=0$ for complex values
(or real ones when below the lowest threshold) of $M$.
We have shown on several occasions in the past
that $D$ contains two types of poles,
which can be distinguished by introducing an overall coupling $\lambda$,
and letting the couplings $\lambda g_{n,\nu}$ continuously
approach zero.
Under this procedure we observe:
\begin{enumerate}
\item
Poles that end up at the mass levels of the confinement
spectrum \cite{KAZIMIERZ83p257,ZPC19p275}.
\label{confinement}
\item
Poles that disappear in the lower half of the complex plane,
with the imaginary part of the pole position \cite{ZPC30p615}
approaching minus infinity.
\label{continuum}
\end{enumerate}

The behavior of the type-\ref{confinement} poles
is exactly as expected for the RSE.
Namely, when the system only couples weakly to the meson-meson sector,
one expects narrow widths and small real mass shifts with respect to
the underlying spectrum, like in photon absorption by hydrogen.
Hence, for small coupling there are poles near each of the mass levels
of the confinement spectrum, which, upon further decreasing
the model's overall coupling, move towards these levels.
For increasing overall coupling, these poles move further into
the lower half of the complex invariant-mass plane, so as to end up at
the positions where they are observed as resonances in experiment.

The type-\ref{continuum} poles are nowadays called
{\it dynamically generated poles},
since they stem from the effective meson-meson interactions,
which in the RSE are automatically contained in the denominator $D$.
They do not have a simple relation to the confinement spectrum,
which does not anticipate their existence.
However, since the complete sum of all $s$-channel diagrams
is equivalent to the sum of all $t$- and $u$-channel diagrams, by duality
\cite{PRL26p1400,JPG2pL167}, it is not a real surprise that upon summing
up a certain class of strongly correlated $s$-channel exchanges
\cite{PRD21p772,PRD27p1527,NTTP4},
one effectively also accounts for part of the $t$-channel exchanges.
Consequently, expression~(\ref{generic}) contains both contributions,
viz.\ the infinity of quark-antiquark resonances as well as
the dynamically generated ones.
This way we can easily generate additional scalar meson nonets,
besides the ones stemming from the quark-antiquark spectrum,
with the same set of parameters that fit the charmonium
and bottomonium bound states and resonances.
\cite{ZPC30p615,AIPCP1030p219}.

In Ref.~\cite{HEPPH0412158}, the authors of the {\it Quarkonium Working Group}
\/(QWG) contended that little work has been done
concerning coupled-channel effects on quarkonium spectra
since the original Cornell model \cite{PRL36p500,PRD17p3090}.
However, they must have overlooked a significant part of the literature
on the subject, as a lot of work has been done over the past 30 years
on the important subject of relating scattering and production data
to models for quarkonia spectra. In particular, a nonperturbative expression
for the scattering amplitude, with the correct analyticity properties
and moreover containing in a natural way the quarkonium spectrum,
has been developed
\cite{PRD21p772,PRD27p1527,NTTP4},
thus enabling us to go beyond perturbative calculations
\cite{PRD69p094019}, which may be unreliable for medium
to large couplings.
Furthermore, in Ref.~\cite{AP323p1215}
an amplitude for production was derived
that allows for the comparison of quarkonium models
with modern data of experimental hadronic physics.
Heavy and light quarkonia are explained
within the same formalism \cite{PRD27p1527,ZPC30p615},
with no need to resort to tetraquarks, meson molecules,
or gluonic excitations \cite{AIPCP1030p219}.
Apparent difficulties for other quark models are no obstacles in the RSE,
but have a common and quite natural explanation,
like the mass of the $D_{s0}(2317)$ \cite{PRL91p012003},
and the masses of the $D_{s1}$(2536)
and $D^{\ast}_{s1}$(2463) mesons \cite{EPJC32p493}.
The RSE  predicted the shape of the
$D_{sJ}(2860)$
production cross section \cite{PRL97p202001},
and even reasonably well the masses \cite{PRD21p772} of the tentative new
\cite{EPL85p61002} charmonium states
$\psi (3D)$, $\psi (5S)$, $\psi (4D)$, $\psi (6S)$,
and $\psi (5D)$. Other authors have made significant contributions to
unitarization and coupled-channel effects as well, also ignored by the
QWG \cite{HEPPH0412158}. It lies outside the scope of the present analysis
to review all these results. Suffice it to mention the coupled-channel
analysis of the $D_{s0}^{\ast}(2317)$ by Hwang and Kim \cite{PLB601p137}, who
computed the mass shift of the $1\,{}^{3\!}P_0$ $c\bar{s}$ state in the
Cornell model, but also including the Coulombic part of the confining 
potential in the calculation of the transition amplitude between the
$c\bar{s}$ and the two-meson sector, in contrast with the common practice
to simply neglect this piece \cite{PRL36p500,PRD17p3090,PRD69p094019}. As
a result, they found a mass shift that is a factor 2.6 larger than what
would result from the standard Cornell procedure, thus allowing to obtain
agreement with experiment. One can only speculate about the consequences
for the Cornell predictions concerning charmonium and especially bottomonium,
which are more compact systems than $c\bar{s}$ mesons, so that the importance
of the Coulombic part of the potential is amplified.

In view of all these results, we are confident that, with some more effort,
most meson-meson scattering and production amplitudes
can be reproduced within the RSE \cite{AP324p1620}.
Nevertheless, in the present paper we
follow a different and more cumbersome path,
in approximating expression~(\ref{prodamp})
by a sum of BW resonances.
This has the advantage of a more direct contact
with experimental methods for extracting resonance poles
from production cross sections.
But more importantly, it gives us the opportunity to find out
what extra ingredients, besides those already included in the RSE,
are necessary for a full description of scattering and production data.
Note, however, that bound states below the lowest threshold
as well as the identification of dynamically generated resonances are
lost in this procedure.  Anyhow, before proceeding, we must first discuss
the coupling of open-bottom pairs to $b\bar{b}$ states.

\subsection{Coupling of open-bottom pairs to \bm{b\bar{b}}}

In the form presented in Eq.~(\ref{prodamp}),
the production amplitude consists of a nonresonant term,
given by a Bessel function, and a sum of terms proportional
to matrix elements of the scattering amplitude.
Resonances reside in the latter matrix elements.
Consequently, we have a clear separation of nonresonant
and resonant terms for the production amplitude.
We shall demonstrate in the following that such a separation
is useful.

Expression~(\ref{prodamp}), which was derived in the framework
of the RSE
\cite{NTTP4}, is a direct link between quark dynamics
and quantities observed in experiment.
However, in its present form the RSE scattering amplitude
does not yet fully correspond to the physical reality,
since it does not account very well for
the experimental observation \cite{PLB69p503}
that two-meson channels only couple significantly
to the quark-antiquark propagator in a very limited
range of energies close to threshold.
From harmonic-oscillator (HO) confinement,
it is reasonably understood
why a specific channel does not survive at higher invariant masses.
Namely, it is straightforward to derive \cite{ZPC17p135,ZPC21p291}
that the number of different channels to which the $q\bar{q}$ system couples
grows rapidly for higher excitations.
So for a given probability of quark-pair creation, the fraction that is left
for a specific channel decreases accordingly.

All terms of the production amplitude of Eq.~(\ref{prodamp})
are proportional to three-meson couplings, given by $g_{\alpha\nu}$.
However, the three-meson couplings also enter
the scattering amplitude $T$ itself \cite{AP323p1215}, that is, their
squares. So at higher energies the terms containing the resonances
fall off faster than the nonresonant term,
leading to $q\bar{q}$ resonance peaks that fade away against
background fluctuations of the nonresonant contribution
in production cross sections. In Refs.~\cite{EPL85p61002,ARXIV09044351},
we showed that signs of the $\psi(5S)$ and $\psi(4D)$
charmonium resonances can be observed just above threshold
in the $e^{+}e^{-}\to\Lambda_{c}^{+}\Lambda_{c}^{-}$
cross section measured by the Belle Collaboration \cite{PRL101p172001},
with tentative $\psi(6S)$ and $\psi(5D)$ signals,
at about 400 MeV higher invariant masses,
almost drowned in the background.

Relative couplings are reasonably well determined by the formalism of
Refs.~\cite{ZPC17p135,ZPC21p291}.
However, as we mentioned above,
the results of the experimental observation obtained with
the SLAC/LBL magnetic detector at SPEAR \cite{PLB69p503}
suggest that at higher invariant masses couplings vanish
more rapidly than as follows from the RSE.
The rate at which the couplings decrease still remains to be fully
understood.

By the use of Eq.~(\ref{prodamp})
and the three-meson couplings of Refs.~\cite{ZPC17p135,ZPC21p291},
we may thus understand why it is hard to observe
$q\bar{q}$ resonance peaks in scattering and especially production.
However, from Ref.~\cite{PLB69p503} we conclude that
such a suppression may even be considerably larger,
ranging from a factor $49\pm 25$, determined via the process
$\sigma\left( DD^{\ast}\right)/\sigma\left( D^{\ast}D^{\ast}\right)$,
to a factor $124\pm 100$, obtained from
$\sigma\left( DD\right)/\sigma\left( D^{\ast}D^{\ast}\right)$,
thereby already accounting for the combinatorial factors $PP:PV:VV=1:4:7$
and scaling invariant mass with the RSE HO frequency $\omega$.
Consequently, we must take one step back in order to figure out what
extra ingredient is needed to comply with the experimental evidence.
Here, we focus on two-meson production reactions in $e^{+}e^{-}$ annihilation,
which we assume to take place via the coupling of the $q\bar{q}$ propagator
to the photon.

\begin{figure}[htbp]
\begin{center}
\begin{tabular}{c}
\scalebox{1.0}{\includegraphics{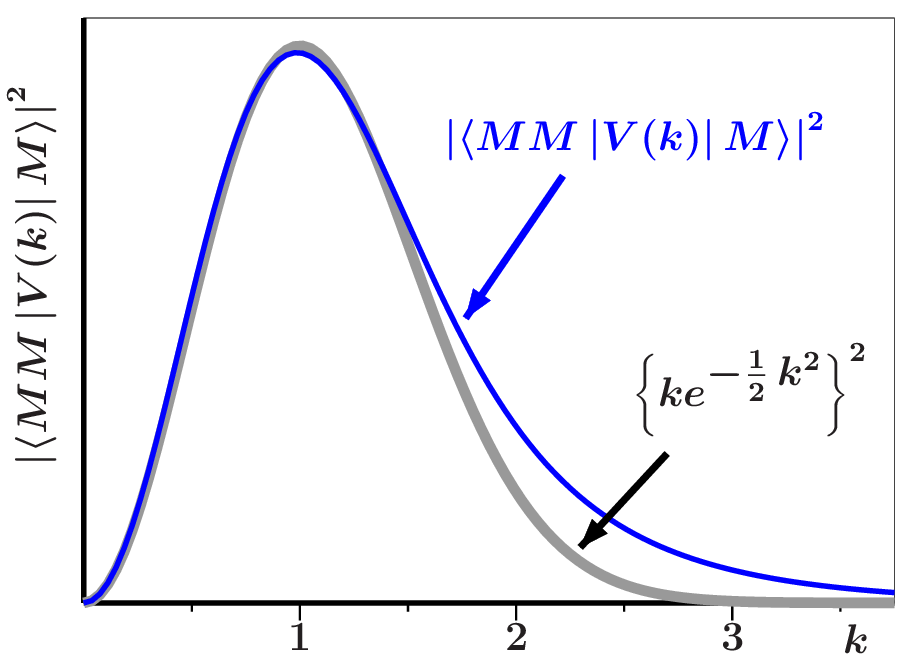}} \\
\end{tabular}\\ [-15pt]
\end{center}
\caption[]{\small The shape of the three-meson vertex strength
$\braket{MM}{V(k)}{M}$ as a function of the relative meson-meson
linear momentum $k$ \lijntje{0}{0}{1},
for $V\to PP$, $V\to PV$ and $V\to VV$
($V=$ vector, $P=$ pseudoscalar)
transitions \cite{ZPC21p291},
assuming HO confinement and $^{3\!}P_{0}$ quark-pair creation,
and its approximation by a Gaussian shape \lijntje{0.6}{0.6}{0.6}.
The matrix element $\braket{MM}{V(k)}{M}$
and the linear-momentum variable $k$,
are dimensionless here.
}
\label{VtoPP}
\end{figure}

In Refs.~\cite{ZPC17p135,ZPC21p291}, a relation between intermesonic distances
and the strength of the three-meson coupling for configuration space was
derived, assuming HO confinement and $^{3\!}P_{0}$ quark-pair creation.
However, since for HO distributions
Fourier transforms are obtained by the simple substitution
of coordinate $x$ by linear momentum $k$,
one readily gets the corresponding relation
between the three-meson coupling and relative linear momentum.
In Fig.~\ref{VtoPP}, we depict the shape
of the three-meson vertex intensity \cite{ZPC21p291}
for the cases to be considered here. In the same figure, we show
that this shape, being an approximation itself,
can be further approximated quite well by a Gaussian shape
\cite{JPG36p075002}.

For $e^{-}e^{+}$ annihilation
into hadrons in the bottomonium region of invariant masses,
we start out with the hypothesis that such processes take place
through the $b\bar{b}$ propagator, and, moreover, that the amplitudes are
dominated by loops of open-bottom mesons.
The latter assumption implies that,
even if the final state contains charmless hadrons,
the line shape of its production cross section is to a large extent
proportional to the line shape of open-charm production.

\section{The BW expansion and its parameters}
\label{parameters}

For each of the open-bottom final states
one has an amplitude of the form given in Eq.~(\ref{prodamp}).
The total amplitude can be represented by a column vector.
Each component of this vector contains the amplitude
for decay into a specific channel.
The intensity of the signal measured in experiment
for one of the channels is proportional
to the square of the modulus of the component
which describes the transition of $b\bar{b}$ to that channel,
e.g.\
\begin{eqnarray}
\lefteqn{
\sigma_{\ell}\left( b\bar{b}\to B\bar{B}\right)
\propto
\abs{a_{\ell}\left( b\bar{b}\to B\bar{B}\right)}^{2}
}
\nonumber\\ [10pt] & &
\propto
\abs{
g_{b\bar{b}\to B\bar{B}}\,
j_{\ell}\left( p_{B\bar{B}}r_{0,\, B\bar{B}}\right)
+\frac{i}{2}\sum_{\nu}\,
h^{(1)}_{\ell}\left( p_{\nu}r_{0,\,\nu}\right)\,
g_{\alpha\nu}\,
T_{\ell}(\nu\to B\bar{B})
}^{2}
\;\;\; ,
\label{procross1}
\end{eqnarray}
where we allow $r_{0}$ to be different for each channel.
The latter ingredient is a generalization of Eq.~(\ref{prodamp}),
But it is useful, as we shall see furtheron.

Each of the scattering amplitudes $T_{\ell}(\nu\to B\bar{B})$
has the form as given in Eq.~(\ref{generic}),
which implies that the pole structure,
given by the zeros of expression~(\ref{denominator}),
is the same for all of them.
As a consequence, the full sum on the righthand side
of Eq.~(\ref{procross1}) has a structure of the form
shown in Eq.~(\ref{generic}).
Hence, we may approximate this full sum by a sum of
BW expressions, according to
\begin{equation}
\frac{i}{2}\sum_{\nu}\,
h^{(1)}_{\ell}\left( p_{\nu}r_{0,\,\nu}\right)\,
g_{\alpha\nu}\,
T_{\ell}(\nu\to B\bar{B})
\to
g_{b\bar{b}\to B\bar{B}}\,\sum_{R}\,
b_{R,\, B\bar{B}}\,
e^\x{i\varphi_{R,\, B\bar{B}}}\,
A_{R}\left(\xrm{BW}\right)
\;\;\; ,
\label{BWexpansion}
\end{equation}
where $R$ runs over all poles in the scattering amplitude,
which are, in principle, infinite in number.
In practice we limit us, of course,
to the number of likely candidates
in the invariant-mass interval under consideration.
Notice that we have suppressed
the relative angular-momentum quantum number $\ell$.

The expansion coefficients that show up in the sum
on the righthand side of Eq.~(\ref{BWexpansion})
are complex numbers.
We have separated the moduli $b_R$ and the phases $\varphi_R$
of these coefficients,
similarly to what is common practice in experimental analysis.
The BW amplitudes are given by
\begin{equation}
A_{R}\left(\xrm{BW}\right)(\sqrt{s})
=\fnd{\omega}{M_{R}-2m_{B}}\,
\fnd{\Gamma_{R}/2}{\sqrt{s}-M_{R}+i\Gamma_{R}/2}
\;\;\; ,
\label{bw}
\end{equation}
where $\sqrt{s}$ and $M_{R}-i\Gamma_{R}/2$ are
the total invariant mass and the pole position
of resonance $R$ in the complex $E$ plane, respectively.
Since the positions of the poles
in the complex invariant-mass plane
are, moreover, the same for all channels,
the BW amplitudes are independent of the channel index.
Furthermore, by the use of Eq.~(\ref{denominator}),
one may deduce that the BW expansion
for the RSE is linear in the masses, and not quadratic
as in the so-called {\it relativistic} \/expressions.
As a consequence, we may find small deviations
from the findings of the \babar\ \/Collaboration
for the resonance pole positions.
The factor $\omega /\left( M_{R}-2m_{B}\right)$
is inserted to account for the invariant-mass dependence
of the $b\bar{b}$ propagator,
which is supposed to be the principal origin of resonances.

For Eq.~(\ref{procross1}) we then get the approximation
\begin{equation}
\sigma_{\ell}\left( b\bar{b}\to B\bar{B}\right)
\propto
\abs{
g_{b\bar{b}\to B\bar{B}}\,
j_{\ell}\left( p_{B\bar{B}}r_{0,\, B\bar{B}}\right)
+
\,\sum_{R}\,
b_{R,\, B\bar{B}}\,
e^\x{i\varphi_{R,\, B\bar{B}}}\,
A_{R}\left(\xrm{BW}\right)
}^{2}
\;\;\; .
\label{procross2}
\end{equation}
At this point we insert the form factor
indicated in Fig.~\ref{VtoPP}
for the coupling of $b\bar{b}$ to $B\bar{B}$,
thereby also introducing some overall constant $\lambda$.
Thus, we obtain the final expression
\begin{eqnarray}
\lefteqn{\sigma_{\ell}\left( b\bar{b}\to B\bar{B}\right) =}
\label{procross3}\\ [10pt] & &
=\lambda^{2}_{B\bar{B}}\, p^{2}_{B\bar{B}}r^{2}_{0,\, B\bar{B}}\,
e^\x{-p^{2}_{B\bar{B}}r^{2}_{0,\, B\bar{B}}}
\abs{
g_{b\bar{b}\to B\bar{B}}\,
j_{\ell}\left( p_{B\bar{B}}r_{0,\, B\bar{B}}\right)
+
\,\sum_{R}\,
b_{R,\, B\bar{B}}\,
e^\x{i\varphi_{R,\, B\bar{B}}}\,
A_{R}\left(\xrm{BW}\right)
}^{2}
.
\nonumber
\end{eqnarray}

The here to be analyzed experimental data for $R_{b}$,
published by the \babar\ \/Collaboration \cite{PRL102p012001},
contain final states of various different open-beauty channels.
We assume that such a signal is proportional to
the sum of the individual cross sections,
by summing them up incoherently, according to
\begin{equation}
R_{b}\propto
\fndrs{1pt}{3s}{-3pt}{4\pi\alpha^{2}}\,
\left\{
\sigma_{\ell}\left( b\bar{b}\to B\bar{B}\right)
+\sigma_{\ell}\left( b\bar{b}\to BB^{\ast}\right)
+\dots
\right\}
\;\;\; .
\label{procross4}
\end{equation}
Note that this expression has enough parameters to fit
a herd of elephants \cite{ChemtechFeb75}.
In the following, we shall discuss these parameters and try to establish
some order among them.

In order to account for channels that open at energies above
the $B_{s}^\ast\bar{B}_{s}^\ast$ threshold,
we introduce a hypothetical seventh channel, called $BX$, with threshold
at $m_{B}+m_{X}$, where $m_{X}$ remains a free parameter. As a result of
our analysis below, we find  for the mass of the open-bottom $X$ meson
an optimum value of \mX\ GeV. This looks
very reasonable in view of the observation
of further $B$-type mesons at energies above 5.72 GeV,
and widths varying from a few tens of MeV to 130 MeV.
We assume here that the $BX$ channel is in a $P$ wave,
although different quantum numbers are certainly possible.
Our $BX$ channel should effectively substitute all possible channels in the
invariant-mass region from about 10.9 GeV upwards.
Its threshold turns out to lie a little bit lower than the
mentioned value of 5.72~GeV, which may account, also in an effective sense,
for the hadronic widths of the open-bottom mesons heavier than the $B_s^\ast$.

\subsection{The 21 phases of the \bm{\Upsilon} resonances}
\label{21phases}

Each of the three resonances considered in this work
has a phase difference with respect to the nonresonant amplitudes
of each of the seven channels involved in our analysis.
So this implies 21 phases in total. If taken as parameters freely adjustable
to the data, this is a sufficiently \cite{ELEFIT} large number 
to fit a little elephant \cite{Babar}.
Fortunately, it is easy to understand what these phases
originate from and therefore how to restrict their arbitrariness.

The bare states of the $b\bar{b}$ spectrum
have a level spacing of $2\omega$
in the HO approximation to the RSE
(HORSE \cite{Clover}). Except for the ground state,
we find two $J^{PC}=1^{--}$ states at each level:
one for which the bottom quarks are in a relative $S$ wave,
and one in a $D$ wave.
Hence, we expect phases to jump by an amount of $2\pi$
over a mass difference of $2\omega$.
This suggests $\pi /\omega$ as the scale for phase differences.

The phase at the mass of resonance $R$
in two-meson channel $A$ must be related
to the mass difference between the channel's threshold
$m_{A1}+m_{A2}$
($m_{A1}$ and $m_{A2}$ are the meson masses)
and the central resonance position $M_{R}$.
If we moreover assume the relation to be linear
to lowest order, then we obtain something of the form
\begin{equation}
\varphi (R,A)=\varphi_{R}+\fnd{m_{A1}+m_{A2}-M_{R}}{\omega}\,\pi
\;\;\; ,
\label{phases}
\end{equation}
where the 3 values for $\varphi_{R}$ would in principle remain
as free parameters. Nevertheless, also these values can be determined,
as we shall explain below.

The central mass positions of resonances do not appear
at the levels of the HO in the HORSE.
Meson loops shift these masses to their physical values.
A comparable situation concerning electromagnetic interactions
exists for the level spectrum of atoms and molecules.
For example, the mass spectrum of the hydrogen atom $H$
can be made visible through electromagnetic interactions.
It appears in the spectrum of $\gamma +H\to\gamma +H$
in the form of narrow absorption lines.
However, as one learns in atomic-physics courses,
the invariant masses of such lines do not correspond
to the exact mass values of the bare hydrogen levels.
They appear at slightly shifted masses
and, moreover, with a certain line width,
both of which can be calculated using perturbative methods.
In the case of strong interactions, perturbative theory is of little
use, whatever beautifull Lagrangians may serve as a starting point.
Nevertheless, one must count with sizable mass shifts and line widths,
which are indeed observed in experiment.
Here, we assume the existence of a bare $b\bar{b}$ spectrum $E_{n\ell}$
($n$ for radial and $\ell$ for angular excitations)
given by the HORSE, reading \cite{PRD21p772,AP324p1620}
\begin{equation}
E_{n\ell}=2m_{b}+\omega\left( 2n+\ell +
\fndrs{-3pt}{\xrm{\scriptsize 3}}{4pt}{\xrm{\scriptsize 2}}\right)
\;\;\; .
\label{Regge}
\end{equation}
The mass shifts and line widths are determined in the RSE
nonperturbatively.

The referred mass shifts cause further phases,
which are opposite in sign.
Now, in general, one does not know very much about the levels
of the bare states.
But in the HORSE they can be determined by using the parameters
optimized and fixed in Ref.~\cite{PRD27p1527}.
For the reference mass we can take any of the states
of the $J^{PC}=1^{--}$ HO spectrum,
since, in the linear approximation,
their phases differ by steps of exactly $2\pi$.
We shall choose here the $3D$ bare-state mass
$M_\xrm{\scriptsize bare}(3D)$,
because it comes out around the mass center
of the resonances to be analyzed.
With the values $\omega =0.190$ GeV
and $m_{b}=4.724$ GeV \cite{PRD27p1527},
and also using Eq.~(\ref{Regge}), we find
\begin{equation}\displaystyle
M_\xrm{\scriptsize bare}(3D)
=2\,m_{b}+\frac{15}{2}\,\omega
=10.873\;\;\xrm{GeV}
\;\;\; ,
\label{bare3D}
\end{equation}
while for $\varphi_{R}$ we obtain
\begin{equation}
\varphi_{R}=\fnd{M_\xrm{\scriptsize bare}(3D)-M_{R}}{\omega}\,\pi
\;\;\; .
\label{restphases}
\end{equation}
Consequently, all 21 phases can be determined,
without any freedom,
by the use of the values of two well-established parameters
for the HORSE.

So how about fitting then the little elephant
\cite{PRL102p012001} of \babar?
Well, the reader should not worry,
there are still plently of parameters left so far.

\subsection{The 21 moduli of the \bm{\Upsilon} resonances}
\label{21moduli}

Each of the three resonances $R$ considered in this work
couples to each of the seven channels $A$
involved in our analysis.
So this implies 21 moduli $b_{R,A}$ in total.
However, if we carefully study the expressions
in Eqs.~(\ref{procross1},\ref{BWexpansion}),
we observe that the BW expansion
stems from a weighted sum over elastic and inelastic matrix elements
of the scattering amplitude.
To leading order, we may assume that the elastic matrix element
dominates the expression.
As a consequence, $b_{R,A}$ is,
to leading order, proportional to $g_{R,A}$,
which allows us to interrelate the moduli,
since the couplings of bare states
to the open-bottom meson-meson channels
just involve combinatorial factors.
Explicitly, writing $B\bar{B}$ for $\abs{b_{R,B\bar{B}}}^{2}$
and similarly for the other channels,
one has for the considered cases
\begin{equation}
B\bar{B}\, :\,
BB^{\ast}\, :\,
B^{\ast}\bar{B}^{\ast}\, :\,
B_{s}\bar{B}_{s}\, :\,
B_{s}B_{s}^{\ast}\, :\,
B_{s}^{\ast}\bar{B}_{s}^{\ast}\, =\,
2\, :\,
8\, :\,
14\, :\,
1\, :\,
4\, :\,
7
\;\;\; .
\label{combinatorics}
\end{equation}

This way, we may limit the number of free parameters to
the moduli $b_{R,B\bar{B}}$
of the three resonances to the $B\bar{B}$ channel.
All the others can be determined
via the combinatorial factors of Eq.~(\ref{combinatorics}).
The seventh channel, viz.\ $B+X$, is treated separately.
Since we assume this channel to be in a $P$ wave,
we may estimate its branching fraction by inspecting
all possible $P$-wave channels with one $B$ meson plus something else.
We find that the total intensity of such channels
is twice as large as the intensity for $B\bar{B}$.
Thus, we use for $B+X$ a combinatorial factor
$\sqrt{2}$.

Furthermore, in the HORSE we may also determine the relative intensities
for $4S$, $3D$ and $5S$ to $B\bar{B}$
\cite{ZPC17p135,ZPC21p291}, obtaining
\begin{equation}
b(4S,B\bar{B})\, :\, b(3D,B\bar{B})\, :\, b(5S,B\bar{B})\, =\,
1\, :\,\frac{2}{\sqrt{3}}\, :\,\frac{\sqrt{11}}{6}
\;\;\; .
\label{4S3D5S}
\end{equation}
Consequently, we are left with one free parameter only, namely
$b(4S,B\bar{B})$, for the 21 moduli pertaining to the three
$\Upsilon$ resonances.
Moreover, the mere fact that
the combinatorial factors of Eq.~(\ref{4S3D5S})
lead to a correct description of the data,
as one may observe in Fig.~\ref{babarups},
lends further support to the assignments
proposed in this paper for the three resonances
at \CivS, \CiiiD, and \CvS\ GeV.

\subsection{The 7 couplings of the meson-meson channels  to \bm{b\bar{b}}}

What is so special about the $b\bar{b}$ case
that cannot be observed equally clearly for $c\bar{c}$,
and even less so in the light quark sector,
is the fact that the lightest thresholds
are separated in one set for open bottom accompanied by light quarks, viz.\
$(b\bar{u})(u\bar{b})$ and $(b\bar{d})(d\bar{b})$,
and another set for open bottom with strange quarks:
$(b\bar{s})(s\bar{b})$.
This very special situation opens a window for a detailed study of
the production of quarks and antiquarks
in electron-positron annihilation.

In the HORSE, the $B\bar{B}$ threshold at 10.56 GeV lies
far above the HO ground state at 9.72 GeV
and the first radial excitation at 10.10 GeV,
and also above the second radial excitation at 10.48 GeV.
Note that the radial level spacings are equal
to $2\omega =0.38$~GeV in the HORSE.
The three-meson vertex intensities for these HO levels
can be determined with the formalism developed in
Ref.~\cite{ZPC21p291}.
The total $B\bar{B}$ branching fraction is found to be
roughly 2.8\% at the HO ground state,
but only about 0.4\% at the second radial HO excitation.
The main competition for $B\bar{B}$ formation
above the $B\bar{B}$ threshold stems from the $BB^{\ast}$ channel.
This competing channel will start to dominate
for invariant masses closer to the $BB^{\ast}$ threshold at 10.60 GeV.
Above the $BB^{\ast}$ threshold, we suppose
that $B\bar{B}$ formation rapidly vanishes.

\begin{figure}[htbp]
\begin{center}
\begin{tabular}{c}
\scalebox{0.7}{\includegraphics{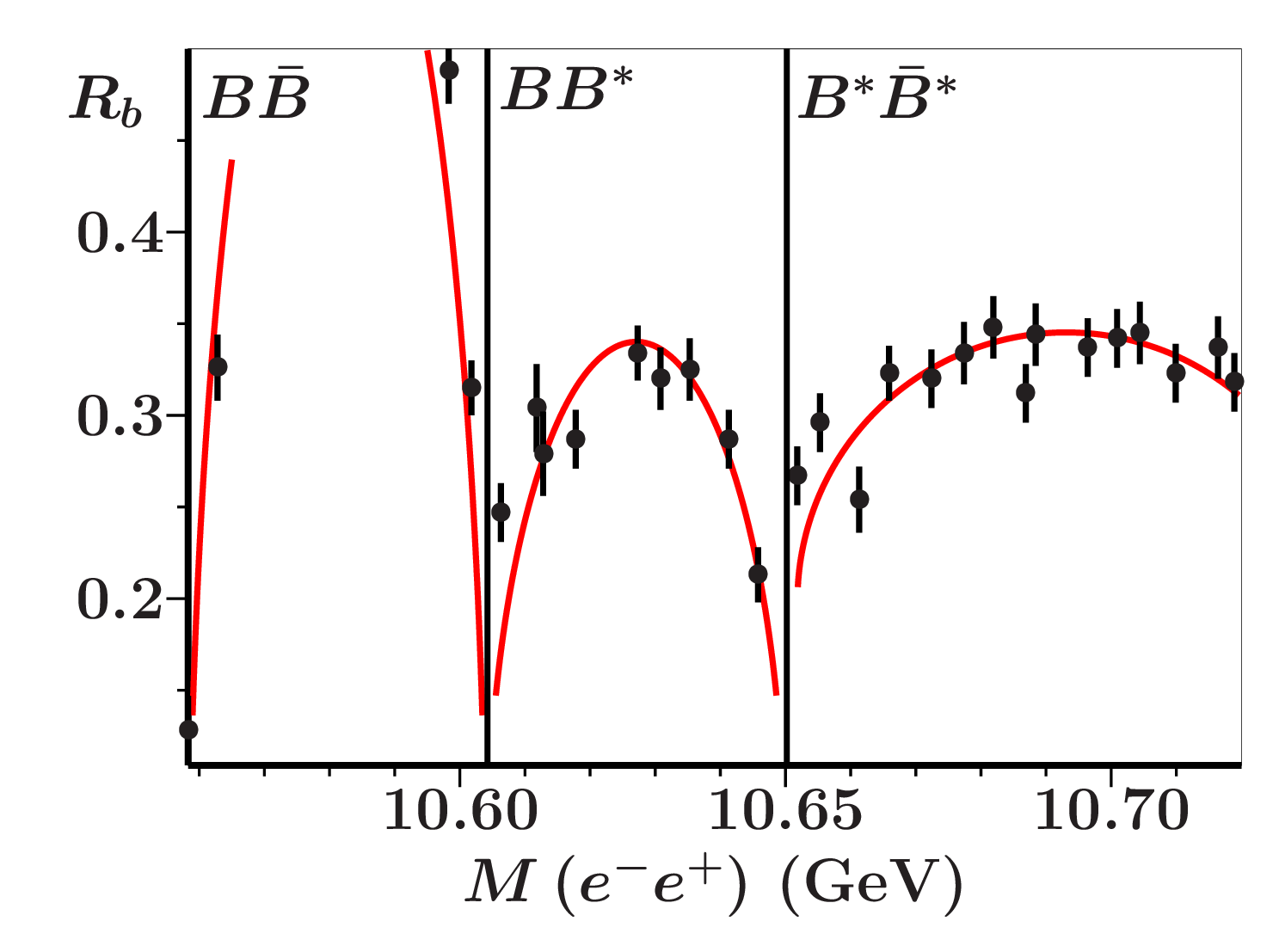}}\\ [-15pt]
\end{tabular}
\end{center}
\caption[]{\small
Experimental data for the process $e^{+}e^{-}\to b\bar{b}$
measured by the \babar\ \/Collaboration \cite{PRL102p012001}.
The vertical lines indicate the $BB^{\ast}$ and $B^{\ast}B^{\ast}$
thresholds, as indicated in the figure.
The eye-guiding lines reflect our interpretation
of the data, and do not represent fits.}
\label{morebabar}
\end{figure}

The above implies that at the $BB^{\ast}$ threshold
the coupling of $B\bar{B}$ to the $b\bar{b}$ propagator
must have decreased significantly.
This is indeed confirmed by Fig.~\ref{morebabar},
in which we show data for the process $e^{+}e^{-}\to b\bar{b}$,
measured and analysed by the \babar\ \/Collaboration \cite{PRL102p012001}.
As also remarked in their paper,
the large statistics and the small energy steps of the
scan make it possible to clearly observe the two dips
at the opening of the thresholds corresponding
to the $B\bar{B}^{\ast}+\bar{B}B^{\ast}$
and $B^{\ast}\bar{B}^{\ast}$ channels.
We have cut off the huge peak at 10.58 GeV,
in order to concentrate better on
the details of the other two enhancements,
at 10.63 GeV and 10.69 GeV, respectively.
Near the $BB^{\ast}$ threshold, we thus observe that
the $B\bar{B}$ signal rapidly
vanishes for increasing invariant mass,
whereas the $BB^{\ast}$ signal behaves the opposite way, i.e.,
it grows fast for increasing invariant mass
just above threshold.
At the $B^{\ast}B^{\ast}$ threshold, this phenomenon is repeated,
now with respect to the $BB^{\ast}$ signal.

So we conclude that $B\bar{B}$ production in $e^{+}e^{-}$ annihilation
can mainly be observed within the invariant-mass window
delimited by the $B\bar{B}$ and $BB^{\ast}$ thresholds.
Similarly, $BB^{\ast}$ production has an equally wide window formed
by the $BB^{\ast}$ and $B^{\ast}B^{\ast}$ thresholds.
The window for $B^{\ast}B^{\ast}$ production is somewhat wider,
since the next threshold concerns the $B_{s}B_{s}$ channel,
which lies considerably higher. Therefore, the enhancement peaking at
about 10.69~GeV is broader than the ones at 10.58 and 10.63~GeV.

\begin{figure}[htbp]
\begin{center}
\begin{tabular}{c}
\scalebox{1.0}{\includegraphics{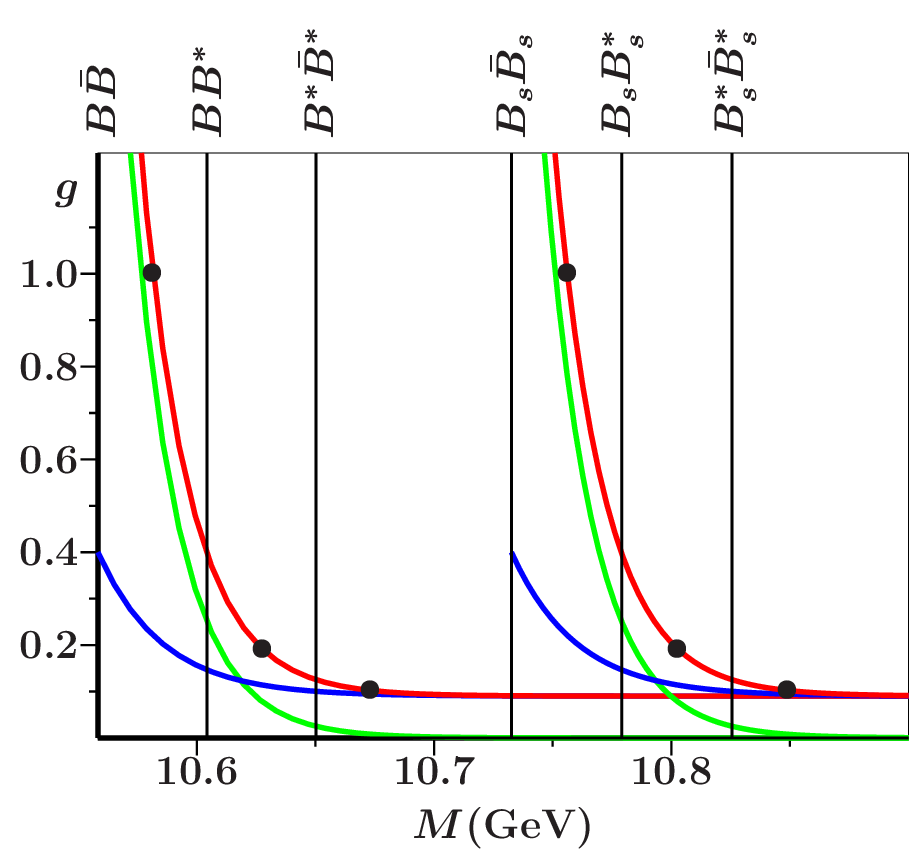}} \\
\end{tabular}\\ [-15pt]
\end{center}
\caption[]{\small
Values of the various parameters $g$ ($\bullet$) used in this work.
From left to right:
$g\left( B\bar{B}\right)$,
$g\left( B\bar{B}^{\ast}\right)$,
$g\left( B^{\ast}\bar{B}^{\ast}\right)$,
$g\left( B_{s}\bar{B}_{s}\right)$,
$g\left( B_{s}\bar{B}_{s}^{\ast}\right)$,
and $g\left( B_{s}^{\ast}\bar{B}_{s}^{\ast}\right)$.
The values are given in the centers of the corresponding
invariant-mass intervals.
The couplings in
the $(b\bar{s})(s\bar{b})$ sector have been multiplied by a factor 4.
The vertical lines indicate the respective thresholds
(see also Fig.~\ref{babarups}).
See the text for an explanation of the curves.
}
\label{couplings}
\end{figure}

The above observations must somehow be contained
in the coupling constants 
$g_{b\bar{b}\to B\bar{B}}$,
$g_{b\bar{b}\to BB^{\ast}}$, and
$g_{b\bar{b}\to B^{\ast}\bar{B}^{\ast}}$
(see Eq.~(\ref{procross3})),
since these quantities regulate the intensity
of the nonresonant contribution to the total production amplitude.
Indeed, after taking out the combinatorial factors
as given in Eq.~(\ref{combinatorics}),
and by defining
\begin{equation}
g_{b\bar{b}\to B\bar{B}}=2g\left( B\bar{B}\right)
\;\; ,\;\;
g_{b\bar{b}\to BB^{\ast}}=8g\left( BB^{\ast}\right)
\;\; ,\;\;
g_{b\bar{b}\to B^{\ast}\bar{B}^{\ast}}=14g\left( B^{\ast}\bar{B}^{\ast}\right)
\;\;\; ,
\label{residualBB}
\end{equation}
we find that the 3 residual couplings
$g\left( B\bar{B}\right)$,
$g\left( BB^{\ast}\right)$, and
$g\left( B^{\ast}\bar{B}^{\ast}\right)$ are not equal,
as one would naively expect,
but that there remains a strong dependence
on invariant mass.
In the left part of Fig.~\ref{couplings},
we show the values of these 3 residual couplings,
normalized to $g\left( B\bar{B}\right)$.
Following an analogous procedure,
also using the combinatorial factors
given in Eq.~(\ref{combinatorics}),
we determine the remaining 3 residual couplings
for the sector where $s\bar{s}$ quark-pair creation
is supposed to lead to bottom-strange meson pairs.
The latter ones are shown in the right part of Fig.~\ref{couplings}.
Moreover, the values for the couplings
in the $(b\bar{s})(s\bar{b})$ sector
are multiplied by a factor 4, in order to better notice
the common behavior of the couplings to $PP$, $PV$, and $VV$
($V=$ vector, $P=$ pseudoscalar).
In each sector, we have depicted 3 curves.
They are scaled with the size of the window,
namely $m\left( B^{\ast}\right)-m\left( B\right)$,
in the bottom-light sector,
and $m\left( B^{\ast}_{s}\right)-m\left( B_{s}\right)$
in the bottom-strange sector.
We assume that our coupling ``constants'' are actually
a kind of average values for the intensity
of open-bottom pair production
in electron-positron annihilation, that is,
averaged over the relevant interval.
Hence, the curves shown in Fig.~\ref{couplings}
represent the picture we imagine for the actual dependence
of these intensities on invariant mass.

We conclude from the results of Fig.~\ref{couplings}
that apparently two different processes are responsible
for the production of pairs of open-bottom mesons
in electron-positron annihilation, namely
one according to the reaction
\begin{equation}
e^{-}e^{+}\to n\bar{n}\to (n\bar{b})(b\bar{n})
\;\;\; ,
\label{eetonbbn}
\end{equation}
for $n$ representing either up or down quarks,
and the other through the reaction
\begin{equation}
e^{-}e^{+}\to b\bar{b}\to (b\bar{n})(n\bar{b})
\;\;\; ,
\label{eetobnnb}
\end{equation}
with an analogous situation in the $(b\bar{s})(s\bar{b})$ sector,
upon substituting $n$ by $s$.
At higher invariant masses, the reaction (\ref{eetobnnb})
and its equivalent upon substitution of $n$ by $s$
seem to dominate the production of pairs of open-bottom mesons,
whereas near the $B\bar{B}$ and $B_{s}\bar{B}_{s}$
thresholds the reaction (\ref{eetonbbn})
and its  $n\to s$ equivalent
appear to be dominant.
We have indicated the evolution of the couplings,
as we imagine it, 
with respect to invariant mass of the former reactions
by \lijntje{0}{1}{0},
and of the latter ones by \lijntje{0}{0}{1}.
The curves \lijntje{1}{0}{0} represents
the sum of the couplings of the two distinct reactions.

At this stage, the reader may like to be informed on the margin
of freedom for the various couplings.
We have not performed a complete analysis of possible error bars
on the values we used for the theoretical curve of
Fig.~\ref{babarups}.
However, as far as we could deduce,
the margin of freedom is small, possibly largest for
$g\left( B_{s}^{\ast}\bar{B}_{s}^{\ast}\right)$.
But even the latter margin does not exceed a few percent, at most 10\%.
Consequently, the error in the factor 4 for the couplings
of the $(b\bar{s})(s\bar{b})$ sector
with respect to the $(b\bar{n})(n\bar{b})$ sector
does certainly not exceed 10\%
and is most probably just a few percent.
In the present work we have fixed this factor at 4, which
might stem from some underlying symmetry we are currently not aware of.
For invariant masses larger
than the upper limit of Fig.~\ref{couplings},
we expect the residual couplings of the two sectors
to be equal.
This implies that the curve of the $(b\bar{n})(n\bar{b})$ sector
must still decrease by a factor 4
with respect to the curve of the $(b\bar{s})(s\bar{b})$ sector.
But this is not relevant for the present analysis.
In total we are thus left with 4 free coupling parameters:
the 3 couplings in the $(b\bar{n})(n\bar{b})$ sector
and the coupling to the seventh channel, $BX$.
The latter coupling, which represents several channels
that open at higher invariant masses,
cannot be related to the others.
Here, we have used the value
$g\left( BX\right)/g\left( B\bar{B}\right) =\BBfacIiv$
for the theoretical curve of Fig.~\ref{babarups}.

Let us now come back to our interpretation of
the results of Fig.~\ref{couplings}.
Near the $B\bar{B}$ and $B_{s}\bar{B}_{s}$ thresholds,
we concluded that open-bottom pairs
dominantly couple to an initial $n\bar{n}$ and $s\bar{s}$ pair, respectively.
The subsequent  $b\bar{b}$ creation then takes place
via an OZI-allowed strong process.
Only for higher invariant masses the primary creation
of $b\bar{b}$ is the main source for open-bottom pairs
via the ensuing OZI-allowed creation of a light quark pair.
Therefore, we may conclude that so far we have bet
on the wrong HORSE, by assuming that primary $b\bar{b}$ creation
is the dominant process above (and below) the $B\bar{B}$ threshold.

As a consequence of these observations,
we seem to may consider initial $b\bar{b}$ creation
only to be dominant for higher invariant masses.
This is possibly what also should be concluded from the data
for the $R$ ratio, where one does not observe a diminishing
of the creation of lighter quark pairs,
but just an additional pair creation of the heavier
type of flavors.
So it is not at all clear
which of the two processes, viz.\ (\ref{eetonbbn}) or (\ref{eetobnnb}),
will be dominant at which energy.
Here, we conclude that the latter reaction
for open-bottom meson-pair creation only starts dominating
at higher invariant masses.

A discussion on the formation of open-bottom mesons
is here in place.
It has always been assumed that once enough energy is available,
meson pairs can be created.
However, in $p\bar{p}$ annihilation that is not
what has been observed \cite{ZPC67p281}.
Although in this process enough energy is available
for the creastion of pion pairs,
Nature prefers to create larger numbers of pions,
most probably via intermediate resonances.
Hence, there seems to be a stability question at hand.
We conclude from the above observation that
a stable pair of $B\bar{B}$ mesons is more easily produced
via an initial pair of $b$ quarks (Eq.~(\ref{eetobnnb})),
but that at low kinetic energies an initial
pair of nonstrange quarks can also produce $B\bar{B}$ mesons.
Light quark pairs are produced more abundantly,
hence dominate near the $B\bar{B}$ threshold.

The fact that this phenomenon is repeated
at the opening of the $B_{s}\bar{B}_{s}$ threshold
lends further credit to our picture.
It also rebuts the possible criticism that by taking
a much larger value for $g_{b\bar{b}\to B\bar{B}}$
than for the other two couplings,
$g_{b\bar{b}\to BB^{\ast}}$ and
$g_{b\bar{b}\to B^{\ast}\bar{B}^{\ast}}$
we {\em generate} \/the enhancement at 10.58 GeV.
Namely, it would be too much of a coincidence to observe
exactly the same phenomenon in the $s\bar{s}$ sector.
In the next section we shall find that the radii corroborate
the above picture.

\subsection{The 7 values for  \bm{r_{0}}}
\label{r0parameter}

The parameter $r_{0}$,
which was introduced in Ref.~\cite{PRD21p772}
in order to allow for a quantity representing
the average distance of quark-pair creation,
requires renewed attention.
\begin{figure}[htbp]
\begin{center}
\begin{tabular}{c}
\scalebox{1.0}{\includegraphics{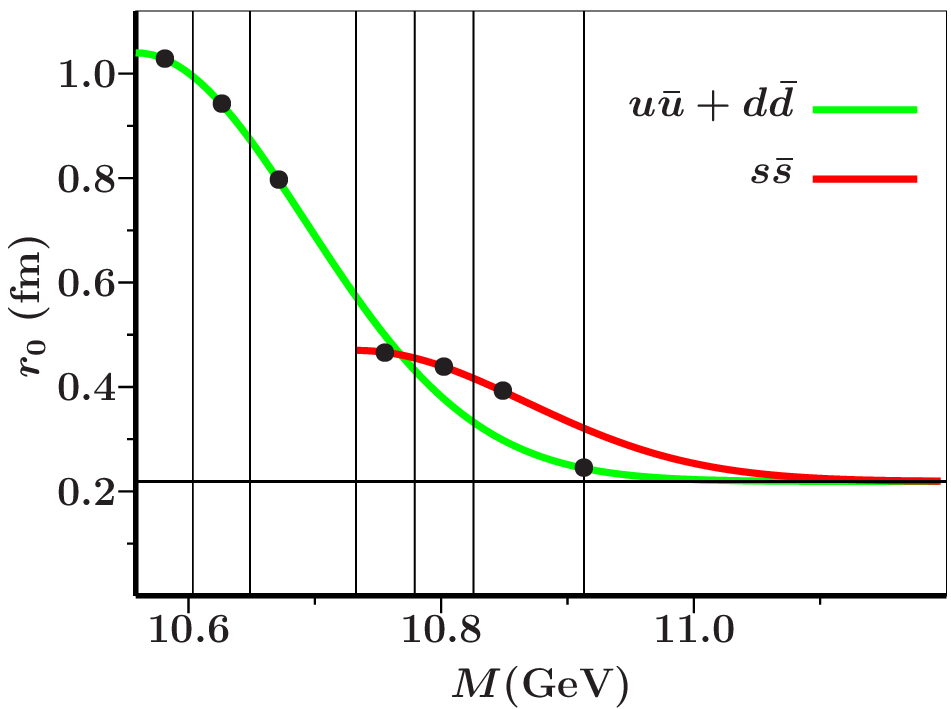}} \\
\end{tabular}\\ [-15pt]
\end{center}
\caption[]{\small
Values of the various parameters $r_{0}$ ($\bullet$) used in this work.
From left to right:
$r_{0}\left( B\bar{B}\right)$,
$r_{0}\left( B\bar{B}^{\ast}\right)$,
$r_{0}\left( B^{\ast}\bar{B}^{\ast}\right)$,
$r_{0}\left( B_{s}\bar{B}_{s}\right)$,
$r_{0}\left( B_{s}\bar{B}_{s}^{\ast}\right)$,
$r_{0}\left( B_{s}^{\ast}\bar{B}_{s}^{\ast}\right)$,
and $r_{0}\left( BX\right)$.
The values are given at the invariant masses where the respective
channels open.
The solid lines, viz.\
\lijntje{0}{1}{0} for $n\bar{n}$
($u\bar{u}$ or $d\bar{d}$)
and \lijntje{1}{0}{0} for $s\bar{s}$ pair creation,
are given by Eqs.~(\ref{stralen},\ref{r0BBbar}),
for $r_{b\bar{b}}=\rbbfm$ fm,
$r_{n\bar{n}} =\rnnfm$ fm
and $r_{s\bar{s}} =\rssfm$ fm.
The vertical lines indicate the respective thresholds
(see also Fig.~\ref{babarups}).
}
\label{radii}
\end{figure}
In Fig.~\ref{radii} we have depicted the values for the various radii
$r_{0}\left( B\bar{B}\right)$,
$r_{0}\left( B\bar{B}^{\ast}\right)$,
$r_{0}\left( B^{\ast}\bar{B}^{\ast}\right)$,
$r_{0}\left( B_{s}\bar{B}_{s}\right)$,
$r_{0}\left( B_{s}\bar{B}_{s}^{\ast}\right)$,
$r_{0}\left( B_{s}^{\ast}\bar{B}_{s}^{\ast}\right)$,
and $r_{0}\left( BX\right)$,
which have been used for the theoretical curve of Fig.~\ref{babarups}.
Just above the $B\bar{B}$ threshold we find values of the order
of 1~fm, corresponding to light quarks,
as expected from the picture for $B\bar{B}$ creation,
discussed in the preceding subsection.
Similarly, we find values of about 0.45 fm
just above the $B_{s}\bar{B}_{s}$ threshold,
corresponding to a system of $s\bar{s}$ pairs.

For higher invariant masses the radii tend towards \rbbfm\ fm,
which corresponds to the size of systems of $b\bar{b}$ pairs.
Hence, the findings for the radii seem to corroborate the picture
that at low kinetic energies
the reaction of Eq.~(\ref{eetonbbn}) dominates,
whereas at higher kinetic energies this reaction cannot provide
the necessary stability for $B\bar{B}$ creation,
so that the reaction of Eq.~(\ref{eetobnnb}) is dominant.

Consequently, we have learned here that our previous approach
to meson-meson production processes, where in the HORSE we only assumed
the reaction of Eq.~(\ref{eetonbbn}) to be relevant,
is not in full agreement with the experimental observations.
Apparently, near the lowest thresholds we must also consider
processes that are dominated by the initial production
of light quark pairs. This very valuable result of the present analysis
should henceforth be incorporated in the HORSE.
For the light quark sector it will not have any influence,
since there is no difference in the reactions
of Eqs.~(\ref{eetonbbn}) and (\ref{eetobnnb})
when $b$ is replaced by $n$.
It may have some consequences for the description of $s\bar{s}$ systems,
but for $c\bar{c}$ and, as we have seen here, for $b\bar{b}$,
inclusion of the light-quark processes will certainly have
some influence near the opening of the lowest lying
open-charm and open-bottom thresholds.
However, for higher invariant masses the results of the HORSE
will not change dramatically.
In particular, the spectra for light and heavy quarkonia
predicted by the HORSE will only change marginally,
as we may conclude from the present analysis.

It is possible to parametrize the curves shown
in Fig.~\ref{radii},
thus reducing the number of effective parameters.
We have opted for the Gaussian expression
\begin{equation}
r_{n\bar{n}}(M)\; =\;
r_{b\bar{b}}+
\left\{ r_{n\bar{n}}
-r_{b\bar{b}}\right\}\,
e^\x{-\left\{ (M-2m_{B\, /\, B_{s}})/\omega\right\}^{2}}
\;\;\; ,
\label{stralen}
\end{equation}
and an analogous one with $n$ replaced by $s$.
For $r_{0}(B\bar{B})$ we have chosen the average value
for $r_{0}(n\bar{n})$ in the window for $B\bar{B}$ production, viz.\
$\Delta (n\bar{n})$, which extends from the $B\bar{B}$ threshold
to the $BB^{\ast}$ threshold,
i.e., $\Delta (n\bar{n})=m_{B^{\ast}}-m_{B}$.
To be more precise, we have just chosen
\begin{equation}
r_{0}(
\left(\begin{array}{c}
B\bar{B}\\ BB^{\ast}\\ B^{\ast}\bar{B}^{\ast}
\end{array}\right)
)=
r_{n\bar{n}}(
M=
\left(\begin{array}{c}
2m_{B}\\ m_{B}+m_{B^{\ast}}\\ 2m_{B^{\ast}}
\end{array}\right)\,
+\fndrs{-3pt}{\xrm{\scriptsize 1}}{4pt}{\xrm{\scriptsize 2}}
\Delta (n\bar{n}))
.
\label{r0BBbar}
\end{equation}
In the $(b\bar{s})(s\bar{b})$ sector
the procedure is analogous, but with the window for $B\bar{B}$ production
replaced by that for $B_{s}\bar{B}_{s}$ production, that is,
$\Delta (s\bar{s})=m_{B^{\ast}_{s}}-m_{B_{s}}$.
For the $BX$ channel we have chosen the value
of our parametrization at the $BX$ threshold.

\subsection{Compilation of the free parameters}

We now absorb the factor $4\pi\alpha^{2}$
of the $R_{b}$ ratio, given in Eq.~(\ref{procross4}),
in the 5 free coupling constants
$b_{\Upsilon(4S),B\bar{B}}$,
$\lambda_{B\bar{B}}$,
$\lambda_{BB^{\ast}}$,
$\lambda_{B^{\ast}\bar{B}^{\ast}}$
and $\lambda_{BX}$ of the 7 contributions.
Our final expression is then given by
\begin{equation}
R_{b}=
\fndrs{1pt}{3s}{-3pt}{\omega^{2}}\,
\left\{
\sigma_{\ell}\left( b\bar{b}\to B\bar{B}\right)
+\sigma_{\ell}\left( b\bar{b}\to BB^{\ast}\right)
+\dots
\right\}
\;\;\; .
\label{scaling}
\end{equation}
The resulting values for the couplings are collected
in Table~\ref{freeparameters}.

We do not include the HORSE frequency $\omega =190$ MeV
in the list of free parameters.
It has been fixed in Ref.~\cite{PRD27p1527}
by carefully studying the average mass splittings
in the light and heavy quarkonia.
In all our work over almost three decades,
we have never changed $\omega$, not even by as little as a tenth of an MeV.
So $\omega$ is something similar to the pion decay constant in other models,
though more stable over the course of time.
Furthermore, $m_{b}=4724$ MeV has also been taken
from Ref.~\cite{PRD27p1527}, and therefore
is not considered a free parameter here.

The parameter $r_{b}$, which could have been determined from
the parameters $\rho_{0}=0.56$ and $m_{b}=4724$ in Ref.~\cite{PRD27p1527},
yielding
\begin{equation}
r_{b}=\fndrs{1pt}{\rho_{0}\hbar c}{-3pt}{\sqrt{\omega m_{b}/2}}
=0.165\;\;\xrm{fm}
\;\;\; ,
\label{rbparameter}
\end{equation}
optimizes here at $r_{b}=0.22$ fm.
This slightly larger value may be due to the different transition potentials
for the $b\bar{b}$ states to the open-bottom meson-meson channels,
as used for the results in Ref.~\cite{PRD27p1527}
and for the formula in Eq.~(\ref{prodamp}), respectively.
Namely, in Ref.~\cite{PRD27p1527} a spatially extended potential was employed,
peaking at a distance given by $\rho_{0}$ and the quark masses,
whereas Eq.~(\ref{prodamp}) results from a potential only acting at $r_{b}$.
The corresponding delta shell does not necessarily
come out at the same position as
the peak of the spatially extended potential.
Therefore, we have considered $r_{b}$ a free parameter here.

In Sec.~\ref{21phases}, we saw that the phases
of the resonances are completely determined
by the parameters $\omega$ and $m_{b}$ of Ref.~\cite{PRD27p1527}.
Hence, there are no free phases.
Furthermore, in Sec.~\ref{21moduli}
we found that all moduli are given by one free parameter, viz.\
$b(4S,B\bar{B})$.
\begin{table}[htbp]
\begin{center}
\begin{tabular}{||l||l||}
\hline\hline & \\ [-7pt]
Parameter(s) & Value(s)\\
& \\ [-7pt]
\hline & \\ [-7pt]
non-interfering background & \Bzero\\ [10pt]
$b_{\Upsilon(4S),B\bar{B}}$ &  \bivS\\ [10pt]
$\lambda_{B\bar{B},BB^{\ast},B^{\ast}\bar{B}^{\ast}}$ &
\BBfacIi, \BBfacIii, \BBfacIiii\\ [10pt]
$m_{X}$, $\lambda_{BX}$ & \mX\ GeV, \BBfacIiv\\ [10pt]
$r_{n\bar{n},s\bar{s},b\bar{b}}$ & \rnnfm\ fm, \rssfm\ fm, \rbbfm\ fm\\ [10pt]
$M_{\Upsilon(4S)}$, $\Gamma_{\Upsilon(4S)}$ &
\CivS\ GeV, \WivS\ MeV\\ [10pt]
$M_{\Upsilon(3D)}$, $\Gamma_{\Upsilon(3D)}$ &
\CiiiD\ GeV, \WiiiD\ MeV\\ [10pt]
$M_{\Upsilon(5S)}$, $\Gamma_{\Upsilon(5S)}$ &
\CvS\ GeV, \WvS\ MeV\\ [10pt]
\hline\hline
\end{tabular}
\end{center}
\caption[]{\small
Parameters used for the theoretical curve of
Fig.~\ref{babarups}.
}
\label{freeparameters}
\end{table}

For the data of the \babar\ \/Collaboration \cite{PRL102p012001},
which amount to the $R_{b}$ ratio for all $e^{+}e^{-}$ annihilation processes
containing $b$ quarks, we assume a background of $R_{b}=\Bzero$, to account for
those reactions that do not contain open-bottom pairs.

A possible criticism that 16 parameters are enough to fit
whatever line shape would not be fair.
In an experimental analysis of the here considered data set,
one would use $3\times 4=12$ BW parameters
for the three resonances,
at least one parametere for the non-interferening background,
and another two parameters for a nonresonant background interfering
with the resonances, thus totaling a minimum number of 15 parameters.
However, such an analysis would certainly not be capable of describing the
$B\bar{B}$ enhancement at 10.580 GeV.
Were the latter structure also described by a BW, the experimental analysis
would need as many as 19 paramenters.
So it seems reasonable to conclude that our excellent description of the
data in Fig.~\ref{babarups} is a good and reliable result.

\section{Results}
\label{Results}

The higher excitations of the bottomonium vector states,
discovered more than two decades ago,
are still today a puzzling topic of intensive research.
In Refs.~\cite{PRL54p377} and \cite{PRL54p381},
the CUSB and CLEO Collaborations, respectively, presented
the first results for the invariant-mass spectra of the
$R\left(\sigma_\xrm{\scriptsize had}/\sigma_{\mu\mu}\right)$
ratio above the open-bottom threshold.

The data of Ref.~\cite{PRL54p377} were observed with
the CUSB calorimetric detector operating at CESR (Cornell).
The experimental analysis resulted in evidence for structures at
$10577.4\pm 1$ MeV,
$10845\pm 20$ MeV, and
$11.02\pm 0.03$ GeV,
with total hadronic widths of
$25\pm 2.5$ MeV,
$110\pm 15$ MeV, and
$90\pm 20$ MeV, respectively.
Structures at about 10.68 and 11.2 GeV were not included
in the analysis of the CUSB Collaboration.

The data of Ref.~\cite{PRL54p381} were obtained from
the CLEO magnetic detector, also operating at CESR.
The experimental analysis resulted in evidence for structures at
$10577.5\pm 0.7\pm 4$ MeV,
$10684\pm 10\pm 8$ MeV,
$10868\pm 6\pm 5$ MeV, and
$11019\pm 5\pm 5$ MeV,
with total hadronic widths of
$20\pm 2\pm 4$ MeV,
$131\pm 27\pm 23$ MeV,
$112\pm 17\pm 23$ MeV, and
$61\pm 13\pm 22$ MeV, respectively.
A structure at about 11.2 GeV was not included
in the analysis of the CLEO Collaboration.

In Tables~\ref{ups10580}, \ref{ups4S}, \ref{ups3D} and \ref{ups5S}
we compare our results for the three resonances
$\Upsilon (4S)$, $\Upsilon (3D)$ and  $\Upsilon (5S)$
to the values published more than two decades ago by
the CUSB \cite{PRL54p377} and CLEO \cite{PRL54p381} Collaborations,
as well as to the more recent data of the \babar\ \/Collaboration
\cite{PRL102p012001}
and the world averages given by the Particle Data Group
in Ref.~\cite{PLB667p1}.

\subsection{The \bm{B\bar{B}} enhancement at 10.580 GeV}

The enhancement at 10.580 GeV was extensively studied
in a more recent publication
\cite{PRD72p032005} of the \babar\ \/Collaboration.
Data were collected with the \babar\ \/detector at the PEP-II storage ring
of Stanford Linear Accellerator Center.
The experimental analysis yielded
$10579.3\pm 0.4\pm 1.2$ MeV and $20.7\pm 1.6\pm 2.5$ MeV
for the central mass and the total (hadronic) width, respectively,
which is in fair agreement with the above results
of the CUSB and CLEO Collaborations.
In the PDG tables~\cite{PLB667p1}, this enhancement is classified as a
$b\bar{b}$ resonance, under the entry $\Upsilon (10580)$.
However, in view of our results, we do
not believe this enhancement to represent a resonance.
Rather, the enhancement at 10.58 GeV suggests
an accumulation of $B\bar{B}$ pairs in this invariant-mass region.
Therefore, a description in terms of a wave function
with a dominant $B\bar{B}$ component appears to be more adequate
than assuming a pole in the scattering amplitude
due to a supposed underlying $b\bar{b}$ state.

More than two decades ago, it naturally seemed obvious to associate
the huge enhancement just above the $B\bar{B}$ threshold
with the $\Upsilon (4S)$ state.  Namely, the generally quite successful
{\it Relativized quark Model} \/(RQM) by Godfrey and Isgur \cite{PRD32p189},
employing the usual funnel-type confining potential, predicted
the $\Upsilon (4S)$ state at 10.63 GeV, i.e., just 50 MeV too high. In view
of the unpredictable threshold effects of the open-bottom decay channels
at the time, this was a rather accurate prediction.
However, the extensive tables of Godfrey and Isgur also contain
an $\Upsilon (3D)$ state at 10.70 GeV.
Why this state was not associated with the resonance at 10.684 GeV
observed by the CLEO Collaboration \cite{PRL54p381},
rather than classifying it as a presumable $b\bar{b}g$
hybrid, is a mystery to us. We conclude here that the {\em latter}
\/state probably is the $\Upsilon (4S)$ resonance.
Table~\ref{ups10580} gives a compilation of the various results
for the $B\bar{B}$ enhancement.
\begin{table}[htbp]
\begin{center}
\begin{tabular}{||c||ccc||}
\hline\hline & & & \\ [-7pt]
Collaboration & Interpretation & Mass (GeV) & Width (MeV)\\
& & & \\ [-7pt]
\hline & & & \\ [-7pt]
RSE (this work) & nonresonant & -- & --\\ [10pt]
CUSB \cite{PRL54p377} & $\Upsilon (4S)$ &
$10.5774\pm 0.001$ & $25\pm 2.5$\\ [10pt]
CLEO \cite{PRL54p381} & $\Upsilon (4S)$ &
$10.5775\pm 0.0007\pm 0.004$ & $20\pm 2\pm 4$\\ [10pt]
\babar\ \/\cite{PRD72p032005,PRL102p012001} & $\Upsilon (10580)$ &
$10.5793\pm 0.0004\pm 0.0012$ & $20.7\pm 1.6\pm 2.5$\\ [10pt]
PDG \cite{PLB667p1} & $\Upsilon (4S)$ &
& $20.5\pm 2.5$\\
\hline\hline
\end{tabular}
\end{center}
\caption[]{\small
Comparison of masses and widths
for the enhancement at 10.58 GeV.
}
\label{ups10580}
\end{table}

\subsection{The \bm{\Upsilon (4S)} resonance}

In Ref.~\cite{PRD21p772}, we found 10.77 GeV
for the real part of the 
$4\,{}^{3\!}S_{1}$ $b\bar{b}$ resonance pole,
in a multichannel calculation
for scattering of open-bottom mesons
This statement is actually not entirely correct,
since $^{3\!}S_1$ and $^{3\!}D_1$ states mix,
so that any resonance contains mixtures of all possible
$^{3\!}S_1$ and $^{3\!}D_1$ states.
Nevertheless, in Ref.~\cite{ZPC19p275} this mixing phenomenon was
studied for charmonium, concluding that mixings are not very large.
So we may say that the resonance at about 10.77 GeV
is dominantly $4\,{}^{3\!}S_{1}$.

The bare spectrum for the HORSE contains
a degenerate $4\,{}^{3\!}S_{1}$/$3\,{}^{3\!}D_{1}$
mass level at 10.873~GeV (see Eq.~(\ref{bare3D})). This level is then split
into two resonances by the coupling to the decay channels,
resulting in one state that is dominantly $3\,{}^{3\!}D_{1}$, with a small
mass shift for the real part of the resonance pole,
and one dominantly $4^{3\!}S_{1}$ state, shifted quite considerably.
We associate the enhancement at \CivS\ GeV found in this work
with the latter resonance, and the bump at \CiiiD\ GeV with the former one.

\begin{figure}[htbp]
\begin{center}
\begin{tabular}{c}
\scalebox{0.7}{\includegraphics{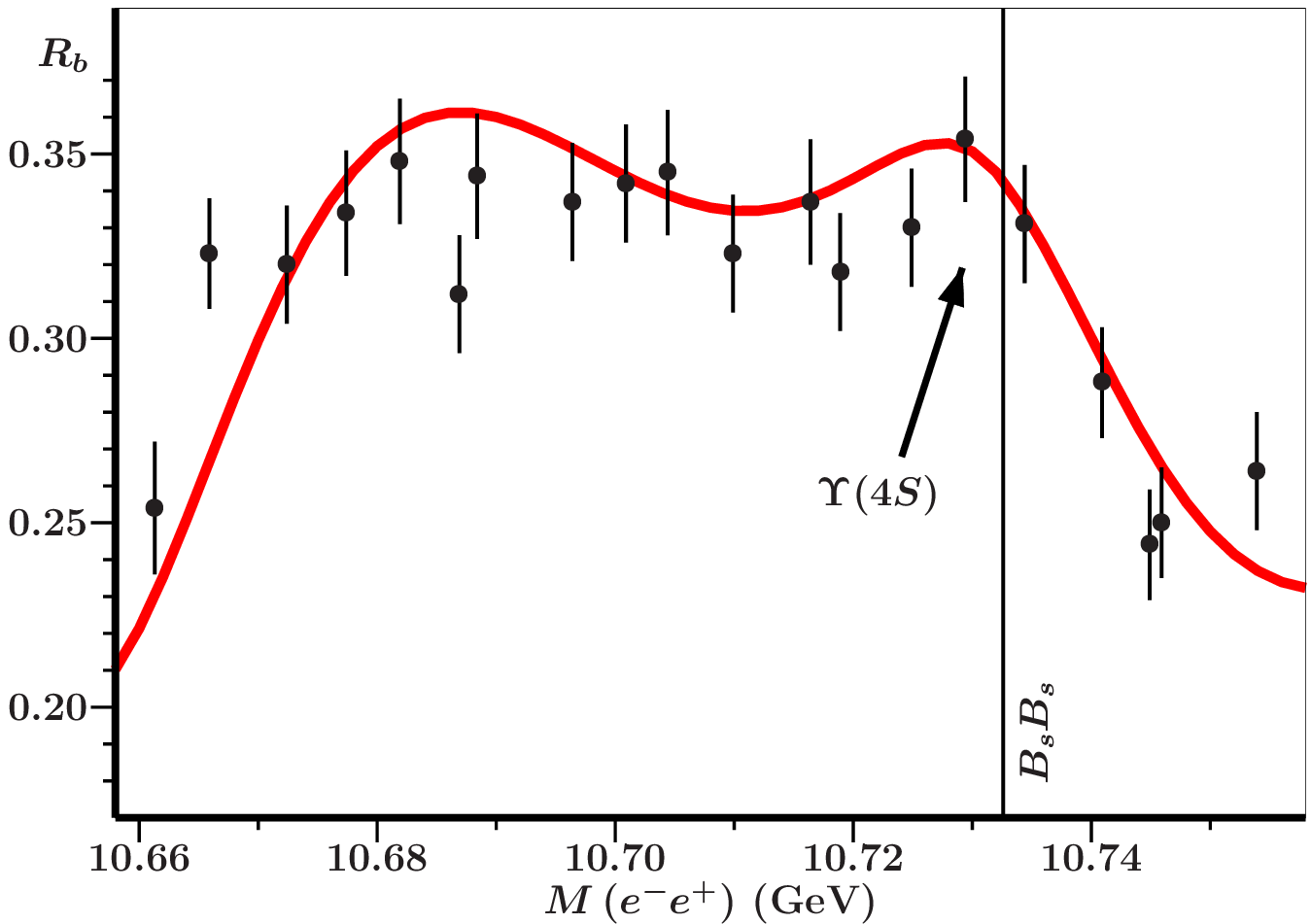}} \\
\end{tabular}\\ [-15pt]
\end{center}
\caption[]{\small
Detail of our result \lijntje{1}{0}{0}
in the $\Upsilon (4S)$ region,
compared to the data ($\bm{\bullet}$)
for hadron production in electron-positron annihilation
published by the \babar\ Collaboration
\cite{PRL102p012001}.
We have baptized this figure the {\it elephit},
as it shows that even with a few parameters one can fit
an elephant (J.~von~Neumann \cite{Nature427p297}).
}
\label{elephit}
\end{figure}

In Fig.~\ref{elephit} we show a detail of Fig.~\ref{babarups}
for the invariant-mas interval 10.658--10.758 GeV,
while in Table.~\ref{ups4S} we compare our result
to the available literature.

\begin{table}[htbp]
\begin{center}
\begin{tabular}{||c||ccc||}
\hline\hline & & & \\ [-7pt]
Collaboration & Identification & Mass (GeV) & Width (MeV)\\
& & & \\ [-7pt]
\hline & & & \\ [-7pt]
RSE (this work) & $\Upsilon (4S)$ & \CivS\ & \WivS\\ [10pt]
CUSB \cite{PRL54p377} & -- & -- & --\\ [10pt]
CLEO \cite{PRL54p381} & $b\bar{b}g$ &
$10.684\pm 0.010\pm 0.008$ & $131\pm 27\pm 23$\\ [10pt]
\babar\ \/\cite{PRL102p012001} & -- & -- & --\\ [10pt]
PDG \cite{PLB667p1} & -- & -- & --\\ [10pt]
\hline\hline
\end{tabular}
\end{center}
\caption[]{\small
Comparison of masses and widths
for the $\Upsilon (4S)$.
}
\label{ups4S}
\end{table}
Our central mass of the $\Upsilon (4S)$
(see Table~\ref{ups4S})
is more than 3 standard deviations
higher than that
of the CLEO Collaboration \cite{PRL54p381}.
On the other hand, for the width we find a much smaller value than theirs.
The reason for either discrepancy is simple.
Namely, we have besides the resonance also a peaking
nonresonant signal, which interferes with the $\Upsilon (4S)$.
Hence, a substantial part of the enhancement does not belong to
the resonance structure, which only corresponds to a very modest signal,
almost coinciding with the $B_{s}\bar{B}_{s}$ threshold.
Its effect is actually best seen just above this threshold,
where it gives rise to the {\it trunk} \/of the elephant-like structure we
baptized {\it elephit} \/(see the caption of Fig.~\ref{elephit}).
As a matter of fact, by moving the central mass position of the $\Upsilon(4S)$
to lower or higher masses, one can wiggle the trunk
of the elephant (J.~von~Neumann \cite{Nature427p297}).
Namely, from Eqs.~(\ref{phases},\ref{restphases}),
also the phase of the resonance depends on its central mass
position.

\subsection{The \bm{\Upsilon (3D)} and \bm{\Upsilon (5S)} resonances}

The identification of the resonance at 10.845 GeV (CUSB) or 10.868 GeV (CLEO)
and the resonance at 11.02 GeV (CUSB) or 11.019 GeV (CLEO)
with the $\Upsilon (5S)$ and $\Upsilon (6S)$ states, respectively,
is in conformity with the RQM \cite{PRD32p189} predictions
at 10.88 GeV and 11.10 GeV, respectively.
\begin{figure}[htbp]
\begin{center}
\begin{tabular}{c}
\scalebox{0.7}{\includegraphics{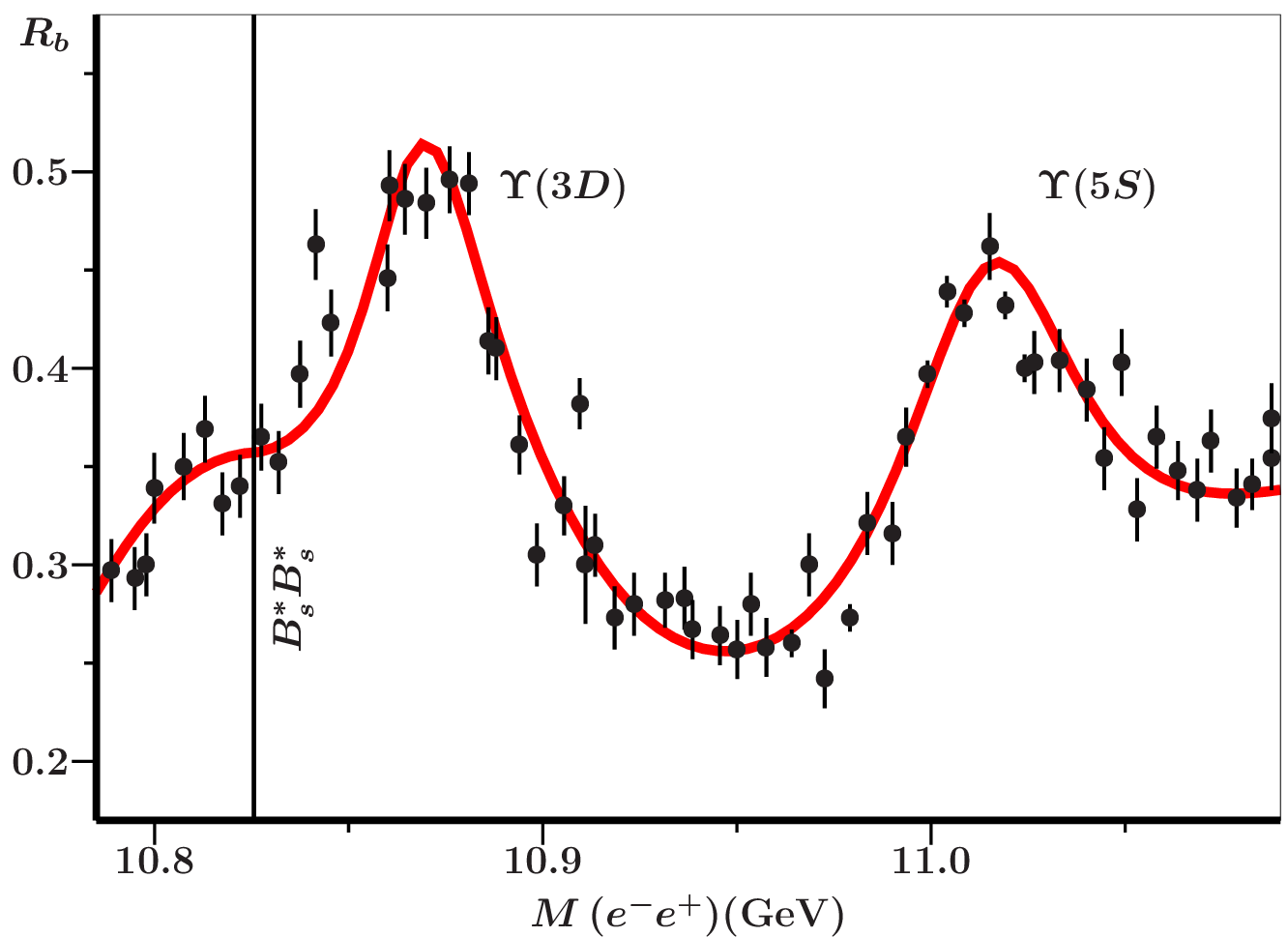}} \\
\end{tabular}\\ [-15pt]
\end{center}
\caption[]{\small
Detail of our result \lijntje{1}{0}{0}
for the $\Upsilon (3D)$ and $\Upsilon (5S)$ resonances,
compared to the data ($\bm{\bullet}$)
for hadron production in electron-positron annihilation
published by the \babar\ \/Collaboration
\cite{PRL102p012001}.
The nonresonant contribution \lijntje{0.6}{0.6}{0.6}
is relatively large in this invariant-mass interval.
One can read the mass of
the $B_{s}^{\ast}\bar{B}_{s}^{\ast}$ threshold from the vertical line.
}
\label{twomore}
\end{figure}
However, we rather identify these resonances with 
the $\Upsilon (3D)$ and $\Upsilon (5S)$ states, respectively.
The reason is that, using a $b$ mass of 4.724 GeV
and an oscillator frequency  of $\omega =0.19$ GeV \cite{PRD27p1527},
the RSE quenched $4S$, $3D$, and $5S$ $b\bar{b}$ states come out at 
10.873 GeV, 10.873 GeV, and 11.253 GeV, respectively.
Unquenching the bare bottomonium spectrum 
by inserting the open-bottom meson loops
\cite{PRD21p772,NTTP4} results in a multichannel scattering amplitude
of the form given in Eq.~(\ref{generic}), which via Eq.~(\ref{prodamp})
also describes production processes. For both types of processes, resonance
poles can be searched for by numerically determining the zeros of $D(\sqrt{s})$
in Eq.~(\ref{generic}). Thus, we find that the $b\bar{b}$ bare states turn
into resonances which for $S$ states have central masses
some 150--200 MeV below the unquenched masses,
whereas the $D$ states undergo mass shifts from a few MeV up to roughly 50~MeV.
These mass shifts largely depend on the precise positions
of the open-bottom thresholds.
In Fig.~\ref{twomore} we show a detail of Fig.~\ref{babarups}
for the invariant-mass interval 10.785--11.090 GeV,
while in Tables~\ref{ups3D} and \ref{ups5S} we compare our findings
to those found in the literature.
\begin{table}[htbp]
\begin{center}
\begin{tabular}{||c||ccc||}
\hline\hline & & & \\ [-7pt]
Collaboration & Identification & Mass (GeV) & Width (MeV)\\
& & & \\ [-7pt]
\hline & & & \\ [-7pt]
RSE (this work) & $\Upsilon (3D)$ & \CiiiD\ & \WiiiD\\ [10pt]
CUSB \cite{PRL54p377} & $\Upsilon (5S)$ &
$10.845\pm 0.020$ & $110\pm 15$\\ [10pt]
CLEO \cite{PRL54p381} & $\Upsilon (5S)$ &
$10.868\pm 0.006\pm 0.005$ & $112\pm 17\pm 23$\\ [10pt]
\babar\ \/\cite{PRL102p012001} & $\Upsilon (10860)$ &
$10.876\pm 0.002$ & $43\pm 4$\\ [10pt]
PDG \cite{PLB667p1} & $\Upsilon (10860)$ &
$10.865\pm 0.008$ & $110\pm 13$\\ [10pt]
\hline\hline
\end{tabular}
\end{center}
\caption[]{\small
Comparison of masses and widths
for the $\Upsilon (3D)$.
}
\label{ups3D}
\end{table}
Our central mass for the $\Upsilon (10860)$
(see Table~\ref{ups3D}) is in good agreement
with those
of the CUSB \cite{PRL54p377},CLEO \cite{PRL54p381},
\babar\ \cite{PRL102p012001} \/Collaborations, and also with the
PDG \cite{PLB667p1} value.
As for the width of the $\Upsilon (10860)$,
we find a much smaller value than CUSB, CLEO, and the PDG,
which is nonetheless in excellent agreement with the very recent
\babar\ \/\cite{PRL102p012001} value. However, on the experimental side
there is a considerable systematic uncertainty with respect to the mass
and width, due to the arbitrariness in the choice of the interfering
background. Let us quote from the \babar\ \/paper \cite{PRL102p012001}
itself (page 7):
\begin{quote}
``The number of states is, a priori,
unknown as are their energy dependencies.
Therefore, a proper coupled channel approach
including the effects of the various thresholds
outlined earlier would be likely to modify the
results obtained from our simple fit.
As an illustration of the systematic uncertainties
arising from the assumptions in our fit,
a simple modification is to replace the flat
nonresonant term
by a threshold function at $\sqrt{s}=2m_{B}$.
This leads to a larger width ($74\pm 4$ MeV)
and a lower mass ($10.869\pm 0.002$ GeV)
for the $\Upsilon (10860)$.''
\end{quote}
The latter value for the width is significantly larger than ours,
but the central mass is now in perfect agreement with our result.

\begin{table}[htbp]
\begin{center}
\begin{tabular}{||c||ccc||}
\hline\hline & & & \\ [-7pt]
Collaboration & Identification & Mass (GeV) & Width (MeV)\\
& & & \\ [-7pt]
\hline & & & \\ [-7pt]
RSE (this work) & $\Upsilon (5S)$ & \CvS\ & \WvS\\ [10pt]
CUSB \cite{PRL54p377} & $\Upsilon (6S)$ &
$11.02\pm 0.03$ & $90\pm 20$\\ [10pt]
CLEO \cite{PRL54p381} & $\Upsilon (6S)$ &
$11.019\pm 0.005$ & $61\pm 13\pm 22$\\ [10pt]
\babar\ \/\cite{PRL102p012001} & $\Upsilon (11020)$ &
$10.996\pm 0.002$ & $37\pm 3$\\ [10pt]
PDG \cite{PLB667p1} & $\Upsilon (11020)$ &
$11.019\pm 0.008$ & $79\pm 16$\\ [10pt]
\hline\hline
\end{tabular}
\end{center}
\caption[]{\small
Comparison of masses and widths
for the $\Upsilon (5S)$.
}
\label{ups5S}
\end{table}
Our central mass of the $\Upsilon (11020)$ (see Table~\ref{ups5S})
agrees well with those of CUSB, CLEO, and the PDG,
but is several standard deviations
larger than the result of \babar.
As for the width of the $\Upsilon (11020)$,
we find a smaller value than CUSB and the PDG,
agree with the CLEO result,
and are six standard deviations away from the \babar\ \/value.
However, as we have just seen, the latter \babar\ \/results depend
quite sensitively on the choice of interfering background.

\subsection{The plateaux in \bm{R_{b}}}

In Ref.~\cite{PRL102p012001}, the \babar\ \/Collaboration
observed two plateaux in $R_{b}$.
The first one comes just below the $\Upsilon (4S)$, and is
depicted in Fig.~\ref{elephit}.
As shown through our theoretical curve,
we do not associate the data with a plateau,
but rather with the back and the shoulders of an elephant.
Also from Figs.~\ref{babarups} and \ref{morebabar}
we seem to learn that neither the nonresonant contribution
nor the resonance have a particularly flat behavior
in the mass region delimited by
the $B^{\ast}\bar{B}^{\ast}$ and $B_{s}\bar{B}_{s}$ thresholds.
As for the nonresonant part, this mass interval
constitutes a window for $B^{\ast}\bar{B}^{\ast}$
production, which signal in part ``carries'' the $\Upsilon (4S)$
resonance.
Furthermore, the tail of the resonance interferes with
the nonresonant contribution, leading to the shallow dip
in between the elephant's back and shoulder.

\begin{figure}[htbp]
\begin{center}
\begin{tabular}{c}
\scalebox{0.7}{\includegraphics{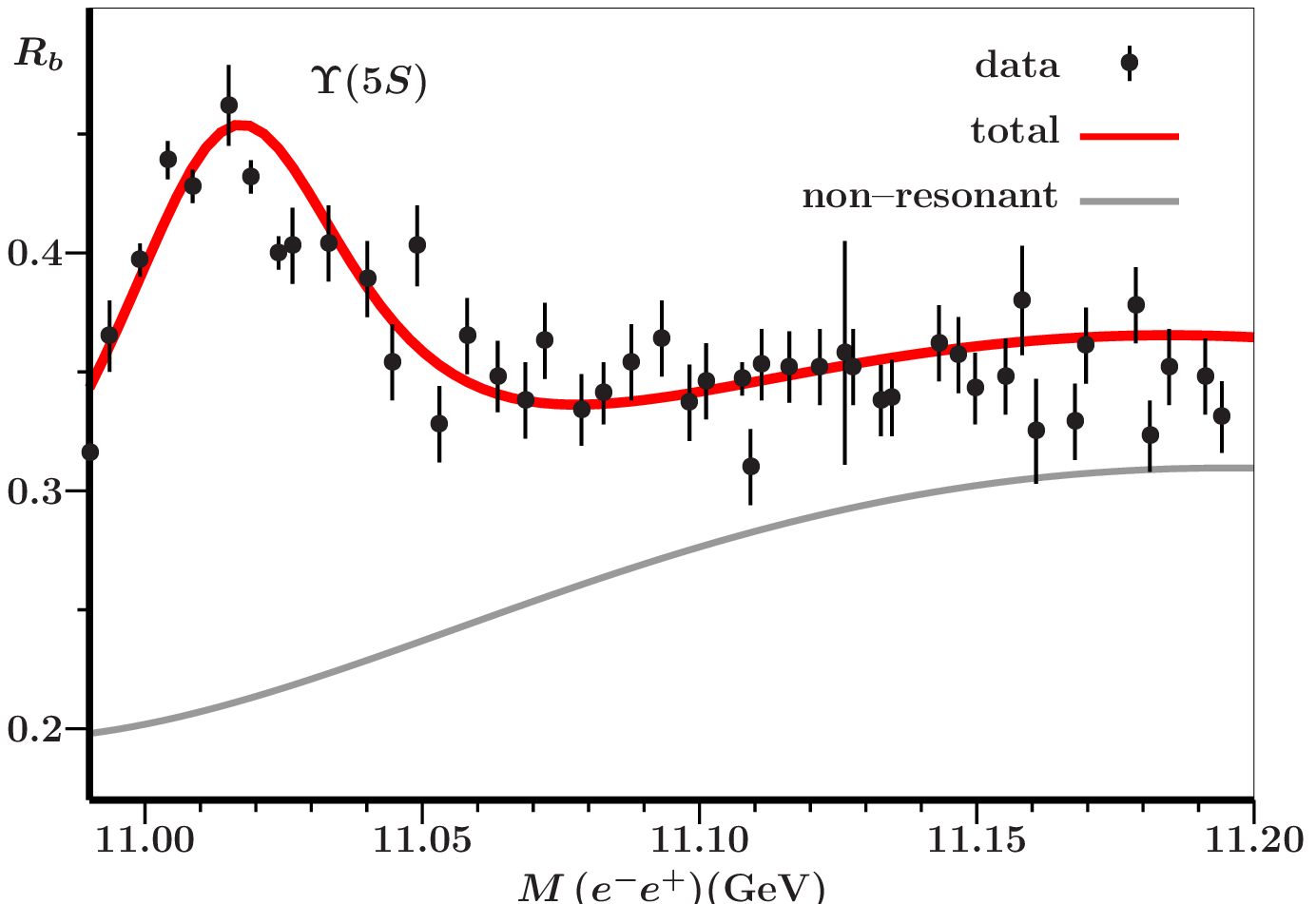}} \\
\end{tabular}\\ [-15pt]
\end{center}
\caption[]{\small
Detail of our result \lijntje{1}{0}{0}
for the $\Upsilon (5S)$ resonance and the ``plateau'',
compared to the data ($\bm{\bullet}$)
for hadron production in electron-positron annihilation
published by the \babar\ \/Collaboration
\cite{PRL102p012001}.
The nonresonant contribution \lijntje{0.6}{0.6}{0.6}
is relatively large and flat in this invariant-mass interval.
}
\label{plateau}
\end{figure}

However, the \babar\ \/Collaboration also points at a second
plateau, above the $\Upsilon (5S)$, which we have depicted
in Fig.~\ref{plateau}.
Here, we indeed observe a flat pattern for $R_{b}$,
which we assume to be the result of a slowly rising
nonresonant contribution
and the tail of the $\Upsilon (5S)$ resonance.
It actually lends further credit to our picture
for open-bottom production in electron-positron annihilation.
Namely, the slowly rising behavior with increasing
invariant mass of the nonresonant contribution
stems from the small value of the $r_{0}$ parameter for $BX$.
This is completely in line with the result
discussed in Sect.~\ref{r0parameter}
and summarized in Fig.~\ref{radii}.

\section{Proposed experiment}
\label{future}

A possible confirmation of the here observed phenomena could come
from an analysis over the same set of data,
but with additional criteria for event selection:
\begin{enumerate}
\item
$B\bar{B}$ events for invariant masses above the $BB^{\ast}$ threshold.
\label{BBbarevents}
\item
Two-meson events containing just one $B$ meson
for invariant masses above the $B^{\ast}\bar{B}^{\ast}$ threshold.
\label{Bevents}
\end{enumerate}
As explained in the preceding sections,
we expect that above the $BB^{\ast}$ threshold
the cross sections of the type-\ref{BBbarevents} events
decrease very fast for increasing invariant masses.
For the type-\ref{Bevents} events, we expect a similar behavior
above the $B^{\ast}\bar{B}^{\ast}$ threshold,
up to invariant masses of about 10.9 GeV.
From 10.9 GeV upwards, we expect the latter cross section
to rise, as with increasing invariant masses
more and more open-bottom $BX$ channels open,
where $X$ represents any excited $B$ meson.
The effective pair-creation radius of this type of channels
is small, of the order of \rbbfm~fm,
which implies that the corresponding signals reach their maxima
roughly 300 MeV above threshold.
A further broadening of the line shapes
may stem from the hadronic widths of the $X$ mesons.
If it turns out to be possible in the analysis of the data sample
to continue up to the $\Lambda_{b}\bar{\Lambda}_{b}$
threshold, one might even observe
a glimpse of the $b\bar{b}$ $4D$ resonance (see below).

As far as spectroscopy is concerned,
just above the $\Lambda_{b}\bar{\Lambda}_{b}$ threshold we expect for
this channel a similar pattern as has been observed
for $\Lambda_{c}\bar{\Lambda}_{c}$ \cite{PRL101p172001}.
However, unlike in the $\Lambda_{c}\bar{\Lambda}_{c}$ case,
where the bare states lie reasonably far away from threshold,
the $\Lambda_{b}\bar{\Lambda}_{b}$ threshold
at $11.2404\pm 0.0022$ GeV \cite{PLB667p1}
comes just below the bare $4D/5S$ state at 11.253 GeV \cite{PRD27p1527},
which could cause a sharp dip in the cross section
at that invariant mass \cite{AP324p1620,EPL85p61002}.
However, it is not
clear to us where the $4D$ resonance will lie precisely.
It may even end up just below the $\Lambda_{b}\bar{\Lambda}_{b}$ threshold,
since the partner $5S$ resonance at \CvS\ GeV is pulled down by as much as
\bareminusCvS\ MeV, which is more than the usual mass shift of $S$-wave
resonances with respect to the corresponding bare states
\cite{PRD21p772}.
However, the $6S$ state, which is probably also pulled down towards
the $\Lambda_{b}\bar{\Lambda}_{b}$ threshold,
will produce a clear signal at roughly 11.4 GeV, so
well below the $\Sigma_{b}\bar{\Sigma}_{b}$ threshold.

\section{Summary and Conclusions}
\label{finalities}

Using a BW approximation to an amplitude
developed for the description of
meson-pair production in electron-positron annihilation
\cite{AP323p1215},
we obtained a satisfactory description of the $R_{b}$ data
very recently published by the \babar\ \/Collaboration
\cite{PRL102p012001}.
The approximation spoils the property of the exact amplitude that
it satisfies Watson's theorem for relating
production and scattering \cite{PR88p1163}.
However, the approximation allowed us to unravel
some of the missing ingredients for a complete description
of hadron production in electron-positron annihilation.
A very important feature here is
that the formalism allows for a well-defined separation
in resonant and nonresonant contributions.

We have found important new ingredients for modeling quarkonia,
in particular $b\bar{b}$ systems.
In the first place, depending on whether the phenomenon that
is most clearly exhibited in Fig.~\ref{couplings}
is correctly interpreted by us,
it seems that open-bottom pair production for small kinetic energies
is dominated by initial light-quark production,
rather than by the heavy $b\bar{b}$ propagator.
An almost exact factor of 4 between the couplings for
$e^{-}e^{+}\to (b\bar{n})(n\bar{b})$ ($n$ for $u$ or $d$)
and the couplings for $e^{-}e^{+}\to (b\bar{s})(s\bar{b})$
remains to be explained.
Secondly, we found that the universal frequency
for strong interactions $\omega$ plays an important role
in relating phases between nonresonant and resonant contributions
to the total production amplitudes
(see Eqs.~(\ref{phases}) and (\ref{restphases})),
and also in interrelating interaction radii
(see Eq.~(\ref{stralen})).
Throughout the present work we have used the value $\omega =0.190$ GeV,
which has been optimized in Ref.~\cite{PRD27p1527} and kept fixed ever since.
Finally, we found that the distributions of Ref.~\cite{ZPC21p291}
for $^{3\!}P_{0}$ quark-pair creation
are useful for the construction of workable form factors
for open-bottom meson-pair production.

Relation~(\ref{4S3D5S}) for the couplings of the three resonances
$\Upsilon (4S)$, $\Upsilon (3D)$ and $\Upsilon (5S)$
to the $b\bar{b}$ propagator
is a clear indication for the here proposed spectroscopic identification
of these resonances.
For the $\Upsilon (4S)$ we have obtained a central resonance mass
of \CivS\ GeV and a width of \WivS\ MeV.

\section*{Acknowledgments}

We are most grateful for the precise measurements
and data analyses of the \babar\ \/Collaboration, 
which made the present analysis possible.
We also thank Dr.\ K.~P.~Khemchandani for many useful discussions.
This work was supported in part by
the \emph{Funda\c{c}\~{a}o para a Ci\^{e}ncia e a Tecnologia}
\/of the \emph{Minist\'{e}rio da Ci\^{e}ncia, Tecnologia e Ensino Superior}
\/of Portugal, under contract CERN/\-FP/\-83502/\-2008.

\newcommand{\pubprt}[4]{{#1 {\bf #2}, #3 (#4)}}
\newcommand{\ertbid}[4]{[Erratum-ibid.~{#1 {\bf #2}, #3 (#4)}]}
\def\AIPCP{AIP Conf.\ Proc.}
\def\AP{Ann.\ Phys.}
\def\EPJC{Eur.\ Phys.\ J.\ C}
\def\EPL{Europhys.\ Lett.}
\def\IJTPGTNO{Int.\ J.\ Theor.\ Phys.\ Group Theor.\ Nonlin.\ Opt.}
\def\JPG{J.\ Phys.\ G}
\def\PLB{Phys.\ Lett.\ B}
\def\PR{Phys.\ Rev.}
\def\PRD{Phys.\ Rev.\ D}
\def\PRL{Phys.\ Rev.\ Lett.}
\def\ZPC{Z.\ Phys.\ C}

\end{document}